\documentclass[10pt,final,journal,twocolumn]{IEEEtran}
\usepackage{hyperref}
\usepackage{ragged2e}

\usepackage{cite}
\usepackage[cmex10]{amsmath}
\usepackage{algorithm}
\usepackage{algorithmic}
\usepackage{stfloats}
\usepackage{multicol,multienum}
\usepackage{multirow}
\usepackage[dvips]{graphicx}
\usepackage{wrapfig}
\usepackage{color}

\usepackage{cite}
\usepackage[cmex10]{amsmath}
\usepackage{amsmath}
\usepackage{algorithm}
\usepackage{algorithmic}
\usepackage{stfloats}
\usepackage{multicol,multienum}
\usepackage{multirow}
\usepackage[dvips]{graphicx}

\usepackage[tight,footnotesize]{subfigure}
\usepackage{wrapfig}
\usepackage{color}

 \usepackage{booktabs}
 \usepackage{multirow}

\usepackage{mathtools}

\usepackage{float}
\usepackage{bm}

\usepackage{pifont}

\usepackage{bbding}
\usepackage{bbding}
\usepackage{pifont}
\usepackage{wasysym}
\usepackage{amssymb}

\usepackage{amsfonts} % Required for mathbb, etc.
\usepackage{graphicx} % Required for resizebox

\begin{document}

\title{\huge Security-Aware Joint Sensing, Communication, and Computing Optimization in Low Altitude Wireless Networks}

\author{
    Jiacheng Wang, Changyuan Zhao, Jialing He, Geng Sun, Weijie Yuan, \\ Dusit Niyato,~\IEEEmembership{Fellow,~IEEE,} Liehuang Zhu,~\IEEEmembership{Senior Member,~IEEE,} Tao Xiang,~\IEEEmembership{Senior Member,~IEEE}
\vspace{-0.3cm}

    \thanks{J.~Wang, C.~Zhao, and D. Niyato are with the College of Computing and Data Science, Nanyang Technological University, Singapore (e-mail: jiacheng.wang@ntu.edu.sg, zhao0441@e.ntu.edu.sg, dniyato@ntu.edu.sg).}

    \thanks{J. He and T. Xiang are with College of Computer Science, Chongqing University, Chongqing 400044, China (e-mail: hejialing@cqu.edu.cn, txiang@cqu.edu.cn).}

    % \thanks{Y. Yang is with National Key Laboratory on Near-Surface Detection, Beijing, 100072, China (e-mail: yaoqi\_yang@yeah.net).}

    \thanks{Geng Sun is with College of Computer Science and Technology, Jilin University, China 130012, (e-mail: sungeng@jlu.edu.cn).}

    % \thanks{Z. Xiong is with the School of Electronics, Electrical Engineering and Computer Science (EEECS), Queen's University Belfast, Belfast, BT7 1NN, U.K. (z.xiong@qub.ac.uk).}

    \thanks{Weijie Yuan is with the School of Automation and Intelligent Manufacturing, Southern University of Science and Technology, Shenzhen 518055, China (e-mail: yuanwj@sustech.edu.cn).}

    \thanks{Liehuang Zhu is with School of Cyberspace Science and Technology, Beijing
Institute of Technology, China (e-mail: liehuangz@bit.edu.cn).}
    % \thanks{Dong In Kim is with the Department of Electrical and Computer Engineering, Sungkyunkwan University, Suwon 16419, South Korea (email: dongin@skku.edu).}

}

\maketitle

\begin{abstract}
As terrestrial resources become increasingly saturated, the research attention is shifting to the low-altitude airspace, with many emerging applications such as urban air taxis and aerial inspection. Low-Altitude Wireless Networks (LAWNs) are the foundation for these applications, with integrated sensing, communications, and computing (ISCC) being one of the core parts of LAWNs. However, the openness of low-altitude airspace exposes communications to security threats, degrading ISCC performance and ultimately compromising the reliability of applications supported by LAWNs. To address these challenges, this paper studies joint performance optimization of ISCC while considering secrecyness of the communications. Specifically, we derive beampattern error, secrecy rate, and age of information (AoI) as performance metrics for sensing, secrecy communication, and computing. Building on these metrics, we formulate a multi-objective optimization problem that balances sensing and computation performance while keeping the probability of communication being detected below a required threshold. We then propose a deep Q-network (DQN)-based multi-objective evolutionary algorithm, which adaptively selects evolutionary operators according to the evolving optimization objectives, thereby leading to more effective solutions. Extensive simulations show that the proposed method achieves a superior balance among sensing accuracy, communication secrecyness, and information freshness compared with baseline algorithms, thereby safeguarding ISCC performance and LAWN-supported low-altitude applications.
\end{abstract}


\section{Introduction}
As ground resources become increasingly scarce, economic activity is shifting from surface infrastructure into low-altitude airspace~\cite{xie2025joint}. Governments and industry are rapidly advancing this domain. For instance, at the government level, the U.S. Federal Aviation Administration is maturing Unmanned Aircraft System Traffic Management (UTM) and Remote ID to coordinate dense low-altitude operations~\footnote{https://www.faa.gov/sites/faa.gov/files/2022-08/UTM\_ConOps\_v2.pdf}. The European Union has enacted the U-space framework to enable high-volume drone activity in designated airspace~\cite{jepsen2024survey}, while China has elevated the low-altitude economy as a growth engine alongside airspace reforms~\cite{wang2025toward}, with early commercial trials of electric vertical take-off and landing (eVTOL) services in tourism and urban mobility scenarios~\cite{zhao2025generative}. In parallel, from commercial perspective, large-scale logistics pilots (e.g., on-demand retail and medical delivery) demonstrate the practicality of drone services across metropolitan and rural regions~\cite{wang2025safeguarding}, while regulators in Asia and Europe continue to harmonize standards through joint workshops and trials. To sustain these emerging use cases at scale, low-altitude wireless networks (LAWN) are essential as the underpinning infrastructure~\cite{yuan2025ground}, providing the sensing, communication, and computation capabilities that interconnect low-altitude vehicles with ground base stations (BSs) and with one another for real-time data exchange, cooperative decision-making, and remote supervision~\cite{he2025satellite}.

Within LAWNs, each low-altitude vehicle functions as an autonomous agent~\cite{zhao2025temporal}, collaborating with peers yet also sense, communicate, compute, and decide under tight latency, spectrum, and energy budgets~\cite{cai2025secure}. This drives the need for integrated sensing, communication and computation (ISCC) as a unifying paradigm~\cite{wen2024survey}. Concretely, integrated sensing and communication has been regarded as a key capability of 6G, enabling communications and sensing to share the same waveform and resources~\cite{wang2024generative}. For example, a low-altitude vehicle can function as both a communication BS and a monostatic radar~\cite{zhang2024cooperative}. It flies over a designated area, transmits a downlink signal to a ground user, and simultaneously uses the same waveform for monostatic sensing~\cite{jing2024isac}. Furthermore, low altitude vehicles can serve as airborne BS platforms equipped with computing resources to deliver edge cloud services on demand~\cite{sun2024joint}. They adaptively perform task scheduling and computation offloading based on their trajectories, link conditions, and operating altitude~\cite{costanzo2020dynamic}. Additionally, the joint design of reconfigurable intelligent surfaces (RIS) and vehicle trajectories can improve line-of-sight (LoS)/non-line-of-sight (NLoS) switching and reduce handover failure rates, thereby enhancing coverage and energy efficiency in complex urban environments~\cite{mei20223d}. These directions are rapidly converging into the ISCC framework that meets the combined requirements of timeliness, energy efficiency, and robustness under dynamic, heterogeneous, and resource-constrained conditions.

However, the openness of low-altitude airspace, introduces pronounced security and privacy risks~\cite{jiang2021secrecy}. High LoS probability and predictable flight paths make wireless links more vulnerable to eavesdropping, jamming, and traffic analysis~\cite{li2024unauthorized}. This directly degrades communication, sensing, and computation performance, which in turn undermines LAWNs stability and reliability and ultimately threatens applications such as air taxis. One commonly used technique for secure transmission is data encryption. By sharing a secret key between two users, the transmitter encrypts the data and the receiver decrypts it~\cite{atoev2019secure}. Without the key, an adversary faces computationally infeasible decryption. Besides that, physical-layer security (PLS) techniques can also be used~\cite{sun2019physical}. By leveraging a stronger legitimate channel than the eavesdropping channel, the eavesdropper cannot reliably decode the message. However, encryption-based schemes require substantial resources for key management and storage~\cite{tan2023secure}, making them impractical for resource-constrained low-altitude aerial vehicles. Moreover, physical-layer security technique cannot fundamentally guarantee the communication security, as once the communication activity is observed by an adversary, the transmitted data becomes vulnerable to decryption. Therefore, secrecy communication, which aims to hide the existence of communication from statistical detectors, is essential for applications in low-altitude scenarios.

Existing studies have investigated ISCC in low-altitude scenarios, but secrecy has not been taken into account. To fill this gap, this paper studies the optimization of sensing, communication, and computation performance of LAWNs under a secrecy communication framework, ensuring effective support for diverse low-altitude applications. We analytically derive the beampattern error, secrecy rate, and data age-of-information (AoI) metrics to characterize the sensing, secrecy communication, and computation performance of LAWN. Building on these metrics, we formulate a multi-objective optimization problem, which aims at improving sensing and computation performance through base-station power allocation while keeping the probability of communication detection within acceptable limits. To solve this problem, we further develop a deep reinforcement learning (DRL)-based multi-objective evolutionary algorithm, which adaptively selects evolutionary operator according to the evolving characteristics of the optimization objective. In summary, the contributions of this paper are as follows.

\begin{itemize}
\item We derive the beampattern error, secrecy rate, and data age-of-information (AoI) metrics, and formulate a joint optimization problem for sensing, secrecy communication, and computation in LAWN. Unlike existing studies, our formulation explicitly incorporates secrecy communication, which is essential for ensuring the stability and security of LAWN.

\item To address the formulated multi-objective optimization problem, we develop a deep reinforcement learning-based multi-objective evolutionary algorithm. This algorithm adaptively adjusts the selection of evolutionary operators according to the dynamics of the optimization objectives, thereby achieving better solutions.

\item We conduct extensive simulations to comprehensively evaluate the performance of the proposed method. Numerical results clearly show that under various practical parameter settings, the proposed scheme can realize a better secrecy communication-oriented ISCC performance than the multi-operator differential evolution and genetic based algorithms.
\end{itemize}

The rest of the paper is arranged as follows. Section II introduces the related works. Then, Section III establishes the system model and formulates the optimization problem. Subsequently, Section IV proposes a DRL-based multi-objective evolutionary algorithm to solve the formulated ISCC optimization problem in LAWNs. On this basis, Section V conducts the simulation experiments to evaluate the performance of the proposed scheme. Finally, Section VI concludes the paper.

%\textcolor{red}{\textbf{Cite all the missed references}}
% \subsection{Motivation and background}
% xxx

% \subsection{Main contributions summarization}

% xxx

% \subsection{Outline of the paper}

% xxx

\section{Related work}

\subsection{Integrated Sensing, Communication and Computation}
Low-altitude wireless networks can support not only data transmission but also high-accuracy sensing and low-latency computation, where ISCC plays a vital role and has been extensively studied by researchers. For instance, the authors in~\cite{xie2024joint} developed a joint sensing, communication, and computation framework for uncrewed aerial vehicle (UAV)-assisted systems and proposes a Dinkelbach’s and successive convex approximation algorithm to efficiently solve the multi-objective optimization problem. Simulations show that the proposed scheme improves the computation–sensing tradeoff region by more than 30\% compared with space-division multiple access and nearly 60\% compared with fully local computing. In~\cite{li2023adaptive}, the authors proposed a cooperative framework for multi-UAV ISCC networks, which optimizes computation offloading and sensing performance. Simulation results show that the framework reduces weighted energy consumption by about 20-30\% compared with existing work, while maintaining stable radar beampattern performance across different user and UAV densities. In addition, the authors in~\cite{deng2024integrated} proposed a multi-UAV ISCC system with adaptive deep neural network (DNN) splitting to jointly optimize sensing, communication, and computation service. The evaluations show that the proposed scheme improves the average sum rate by up to 25-40\% compared with benchmark schemes, while reducing total latency by over 30\% through flexible DNN layer splitting and UAV location optimization.

Beside ISCC performance, trajectory is also a key factor. For instance, the authors in~\cite{zhou2023uav} introduced a UAV-enabled ISCAC framework for IoT that minimizes weighted energy consumption through joint optimization of CPU frequency, sensing power, UE transmit power, and UAV trajectory. Simulations demonstrate that the proposed algorithm reduces overall energy consumption by about 20-35\% compared with relaying-only baselines, while simultaneously achieving higher radar estimation rates. Another work~\cite{chen2024uav} proposed a UAV-assisted ISCC framework with joint optimization of trajectory, beamforming, and offloading strategy to maximize computing efficiency. Evaluations demonstrate that the proposed algorithm outperforms benchmark schemes, enhancing computing efficiency by approximately 20-35\% over maximizing computing rate and minimizing energy consumption, while sustaining stable performance across varying flight durations.

\subsection{Secure Communication}
The authors in~\cite{zhou2019uav} presented a dual-UAV design, which combines UAV-BS trajectory/power with a cooperative jammer to maximize the min secrecy rate. The evaluations show that alternating successive convex approximation yields large gains, which are about 20\% over joint design without jammer. Besides that, authors in~\cite{zhang2020uav} proposed a cooperative jamming approach by letting UAV jammers help the UAV transmitter defend against ground eavesdroppers. They presented a multi-agent deep reinforcement learning approach to maximize the secure capacity by jointly optimizing the trajectory of UAVs, the transmit power from UAV transmitter and the jamming power from the UAV jammers. Simulations show that using three jammers achieves a secure rate of about 4.5 bps/Hz, which is about 2.5 bps/Hz higher than with one jammer. The authors in another work~\cite{wang2021robust} proposed a dual-UAV-enabled secure communication system, in which, 3D mobility, energy harvesting with time-switching, mobile user, and imperfect eavesdropper locations are considered to maximize worst-case secrecy. They decoupled the optimization problem into three subproblems and develop an iterative algorithm to find its suboptimal solution by using the block coordinate descent technique. Simulations show that this method achieved higher secrecy rates than a conventional 2D trajectory scheme. For the UAV-assisted mobile edge computing scenario with eavesdroppers, authors in~\cite{ding2022online} proposed a scheme, which maximizes secure computing efficiency by jointly optimizing users' binary task offloading decisions and UAV resource allocation via deep reinforcement learning and iterative convex optimization. Simulations show that the performance of the proposed method approaches that of the exhaustive search solution. Moreover, the authors in~\cite{tian2023uav} maximized the sum secrecy capacity Bob received by optimizing the transmit power and the UAV’s trajectory jointly while guaranteeing that the normal communication behavior between ground device-to-device pairs is not affected. They proposed a jointly optimizing the transmit power and UAV trajectory algorithm based on the block coordinate descent, which can further improve the secrecy capacity compared with existing works.

From the above analysis, it is clear that, from the ISCC perspective, existing studies have primarily focused on performance optimization while overlooking communication security~\cite{wang2024joint}. In low-altitude airspace, applications such as air taxis impose stringent security requirements, making current ISCC research insufficient to meet practical demands. From the perspective of secrecy communication, prior studies have concentrated mainly on reducing the probability of communication being detected, without considering the multifunctional roles of aerial vehicles and base stations. In response to this research gap, this work focuses on optimizing sensing, communication, and computation in LAWNs under secrecy communication constraints, thereby ensuring robust and effective support for diverse low-altitude services.

\section{System model and problem formulation}
In this section, we build models from low-altitude ISCC perspectives, including sensing, secrecy communication, and computing, and then formulate the optimization objective.

\begin{figure}[!htb]
  \centering
  \includegraphics[width=8cm]{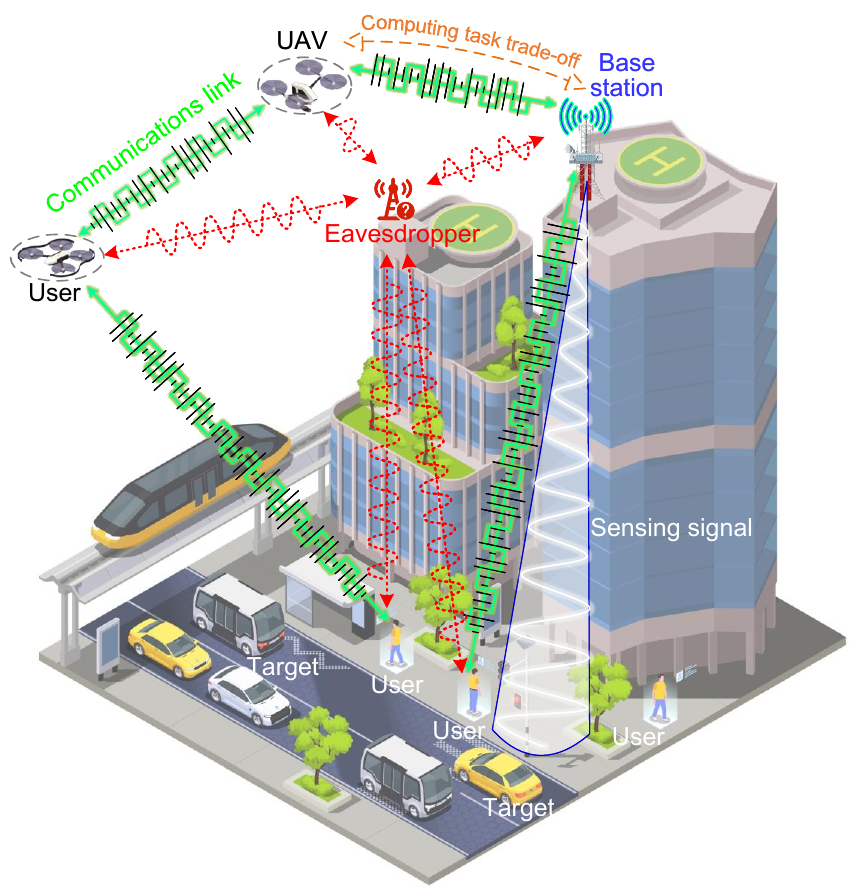}\\
  \caption{System model of the security-aware sensing, communication, and computing in low altitude wireless networks. In the considered LAWNs, the BS provides sensing, communication, and computing services to low-altitude applications such as air taxis in a time-division manner. Given the openness of the low-altitude airspace, LAWNs inject noise into the transmitted communication signals to reduce the probability of being detected by eavesdropper, thereby enhancing the network security.}
  \label{SYSTEMMODEL}
\end{figure}

\subsection{ISCC in LAWNs}
As shown in Fig.~\ref{SYSTEMMODEL}, in the LAWN, an ISCC BS provides sensing, communication and computation services in a time division multiplexing manner \cite{xie2024joint}. Specifically, during the sensing phase, the BS can launch the sensing beam to perceive targets on the ground or in the low-altitude airspace. Besides, the BS can communicate with UAVs and also serve distant users via relay UAVs. However, due to the openness of LAWN, potential eavesdroppers may intercept and monitor legitimate wireless transmissions. To address this threat, the BS employs secrecy communication by embedding artificial noise (AN) into the transmitted signals, thereby degrading the channel quality of eavesdroppers. In addition, the BS can provide computation services for relay UAVs, thereby alleviating the stress regarding the energy consumption and the timeliness data transmission. For clarity, we denote the number of antennas at the BS, relay UAV, mobile user, and eavesdropper as $N_{BS}$, $N_{UAV}$, and $N_{evae}$, respectively, while $N_{user}$ representing the number of users with a single antenna. The total transmit power of the BS, $P_{BS}$, is allocated to three components, including $P_{sens}$ for sensing, $P_{com}$ for communication, and $P_{AN}$ for artificial noise generation. Hence, the transmitted signal of the BS can be expressed as 
% \begin{equation}\label{11111}
% {\bf{x}} = \left\{ \begin{array}{l}
% \sqrt {{P_{com}}} {{\bf{W}}_{\mathop{\rm c}\nolimits} }{{\bf{x}}_{\mathop{\rm c}\nolimits} } + \sqrt {{P_{AN}}} {{\bf{V}}_{\mathop{\rm n}\nolimits} }{{\bf{z}}_{\mathop{\rm n}\nolimits} },\;{\mathop{\rm comm}\nolimits}\rm \& compu \quad mode\\
% {{\bf{W}}_{\mathop{\rm r}\nolimits} }{{\bf{x}}_{\mathop{\rm r}\nolimits} },\;{\mathop{\rm sensing}\nolimits}  \quad \rm mode
% \end{array} \right.,
% \end{equation}
\begin{equation}\label{11111}
{\bf{x}} = \left\{ 
\begin{array}{l@{\;}l}
\sqrt{P_{com}}\,{\bf W}_{c}{\bf x}_{c} + \sqrt{P_{AN}}\,{\bf V}_{n}{\bf z}_{n}, 
& {\mathop{\rm comm}\nolimits}\,\&\,{\rm compu~mode} \\[4pt]
{\bf W}_{r}{\bf x}_{r}, 
& {\mathop{\rm sensing}\nolimits}~{\rm mode}
\end{array}
\right.,
\end{equation}
where ${{\bf{W}}_{\mathop{\rm c}\nolimits} } \in \mathbb{C}{^{{N_{BS}} \times {N_{user}}}}$ is the zero force precoder, ${{\bf{x}}_{\mathop{\rm c}\nolimits} } = {\left[ {{x_{r1,}}{x_{r2,}} \ldots ,{x_{r{N_{user}}}}} \right]^{\mathop{\rm H}\nolimits} } \in \mathbb{C}{^{{N_{user}} \times 1}}$ represents the transmitted symbol vector, which satisfies $\mathbb{E} \left[ {{{\bf{x}}_{\mathop{\rm c}\nolimits} }{{\bf{x}}_{\mathop{\rm c}\nolimits} }^{\mathop{\rm H}\nolimits} } \right] = {{\bf{I}}_{{N_{user}}}}$, ${{\bf{V}}_{\mathop{\rm n}\nolimits} } \in {\cal C}{{\cal N}_{{N_{BS}} \times {N_{UAV}}}}$ means the AN shaping matrix of the BS, and ${{\bf{z}}_{\mathop{\rm n}\nolimits} } \sim {\cal C}{{\cal N}_{{N_{UAV}} \times 1}}\left( {{{\bf{0}}_{{N_{UAV}} \times 1}},\sigma _Z^2{{\bf{I}}_{{N_{UAV}}}}} \right)$ is the AN vector. Moreover, ${{\bf{W}}_{\mathop{\rm r}\nolimits} } \in \mathbb{C}{^{{N_{BS}} \times {N_{BS}}}}$ is the precoder matrix for the sensing signals, ${{\bf{x}}_{\mathop{\rm r}\nolimits} } = {\left[ {{x_{c1,}}{x_{c2,}} \ldots ,{x_{c{N_{BS}}}}} \right]^{\mathop{\rm H}\nolimits} } \in \mathbb{C}{^{{N_{BS}} \times 1}}$ is the vector of the individual sensing signals. On this basis, the covariance matrix of the transmit sensing signal can be calculated as:
\begin{equation}\label{15615}
{{\bf{R}}_{\mathop{\rm r}\nolimits} } = \mathbb{E}\left[ {{{\bf{W}}_{\mathop{\rm r}\nolimits} }{{\bf{x}}_{\mathop{\rm r}\nolimits} }{{\left( {{{\bf{W}}_{\mathop{\rm r}\nolimits} }{{\bf{x}}_{\mathop{\rm r}\nolimits} }} \right)}^{\mathop{\rm H}\nolimits} }} \right] = {{\bf{W}}_{\mathop{\rm r}\nolimits} }{{\bf{W}}_{\mathop{\rm r}\nolimits} }^{\mathop{\rm H}\nolimits} .
\end{equation}

\subsection{Sensing Model}
During the sensing phase, the BS sends out the narrow-band signal, and the target is in the line-of-sight (LoS) situation~\cite{da2023privacy}. The sensing signal in the direction of $\theta $ can be expressed as:
\begin{equation}\label{4785}
{{\bf{y}}_\theta } = {{\bf{a}}^{\mathop{\rm H}\nolimits} }\left( \theta  \right){{\bf{x}}_{\mathop{\rm r}\nolimits} },
\end{equation}
where ${\bf{a}}\left( \theta  \right)$ means the array steering vector. Let $\Delta $ represent the normalized antenna separation, ${\bf{a}}\left( \theta  \right)$ can be further denoted as:
\begin{equation}\label{956562}
{\bf{a}}\left( \theta  \right) = \left[ {1\quad {e^{j2\pi \Delta \sin \left( \theta  \right)}} \ldots {e^{j2\pi \left( {{N_{BS}} - 1} \right)\Delta \sin \left( \theta  \right)}}} \right] \in \mathbb{C}{^{{N_{BS}} \times 1}},
\end{equation}
and the beampattern in the direction of $\theta$ is
% \begin{equation}\label{78951211}
% {{\bf{B}}_{{\mathop{\rm pattern}\nolimits} }} = \mathbb{E}\left[ {{{\left| {{{\bf{y}}_\theta }} \right|}^2}} \right] = \mathbb{E}\left[ {{{\bf{a}}^{\mathop{\rm H}\nolimits} }\left( \theta  \right){{\bf{x}}_{\mathop{\rm r}\nolimits} }{{\bf{x}}_{\mathop{\rm r}\nolimits} }^{\mathop{\rm H}\nolimits} {\bf{a}}\left( \theta  \right)} \right] = {{\bf{a}}^{\mathop{\rm H}\nolimits} }\left( \theta  \right){{\bf{R}}_{\mathop{\rm r}\nolimits} }{\bf{a}}\left( \theta  \right).
% \end{equation}
\begin{equation}\label{78951211}
{{\bf{B}}_{{\mathop{\rm pattern}\nolimits} }}= \mathbb{E}\left[ {{{\bf{a}}^{\mathop{\rm H}\nolimits} }\left( \theta  \right){{\bf{x}}_{\mathop{\rm r}\nolimits} }{{\bf{x}}_{\mathop{\rm r}\nolimits} }^{\mathop{\rm H}\nolimits} {\bf{a}}\left( \theta  \right)} \right] = {{\bf{a}}^{\mathop{\rm H}\nolimits} }\left( \theta  \right){{\bf{R}}_{\mathop{\rm r}\nolimits} }{\bf{a}}\left( \theta  \right).
\end{equation}

Given $P_{sens}$ is the transmission power of the sensing signal, the diagonal elements of ${{{\bf{R}}_{\mathop{\rm r}\nolimits} }}$ is:
\begin{equation}\label{624545}
{\left[ {{{\bf{R}}_{\mathop{\rm r}\nolimits} }} \right]_{i,i}} = {{{P_{sens}}} \mathord{\left/
 {\vphantom {{{P_{sens}}} {{N_{BS}}}}} \right.
 \kern-\nulldelimiterspace} {{N_{BS}}}},i = 1,2, \ldots ,{N_{BS}}.
\end{equation}
Let $\lambda $ represent the scaling factor, $\left( {{\theta _l}} \right)_{l = 1}^L$ denote the sampled angle grids, and $d\left( {{\theta _l}} \right)$ denote the desired beampattern, then the beampattern error can be denoted as
\begin{equation}\label{15646564785}
R{B_{error}} = \frac{1}{L}{\sum\nolimits_{i = 1}^L {\left| {\lambda d\left( {{\theta _l}} \right) - {{\bf{a}}^{\mathop{\rm H}\nolimits} }\left( {{\theta _l}} \right){{\bf{R}}_{\mathop{\rm r}\nolimits} }{\bf{a}}\left( {{\theta _l}} \right)} \right|} ^2}.
\end{equation}
beampattern error is used to measure the sensing performance of the BS, where lower values indicate better sensing performance.
\subsection{secrecy Communication Model}
For the communication, let ${\bf{M}} \in \mathbb{C}{^{{N_{UAV}} \times {N_{BS}}}}$, ${\bf{N}}\in \mathbb{C}{^{{N_{user}} \times {N_{UAV}}}}$, ${\bf{Q}} \in \mathbb{C}{^{{N_{eave}} \times {N_{BS}}}}$ and ${\bf{Z}} \in \mathbb{C}{^{{N_{eave}} \times {N_{UAV}}}}$ be the channel between BS and relay UAV, the channel between relay UAV and mobile users, the channel between BS and the eavesdropper, and the channel between relay UAV and the eavesdropper, respectively. These channels can be uniformly expressed as
\begin{equation}\label{2256146}
{\bf{C}} = {\bf{F}}_C^{{1 \mathord{\left/
 {\vphantom {1 2}} \right.
 \kern-\nulldelimiterspace} 2}}{\bf{\tilde C}}
\end{equation}
where ${\bf{C}} \in \left\{ {{\bf{M}},{\bf{N}},{\bf{Q}},{\bf{Z}}} \right\}$, ${\bf{\tilde C}} \sim {\cal C}{{\cal N}_{s \times t}}\left( {{{\bf{0}}_{s \times t}},{{\bf{I}}_s} \otimes {{\bf{I}}_t}} \right)$ represents the independent and frequency-flat small-scale Rayleigh fading. Here, the path-loss $P_{loss}$ is subject to $ {\left( {{{di{s_{m,n}}} \mathord{\left/
 {\vphantom {{di{s_{m,n}}} {di{s_{ref}}}}} \right.
 \kern-\nulldelimiterspace} {di{s_{ref}}}}} \right)^{{L_e}}}$, where ${di{s_{ref}}}$ is the reference distance, ${di{s_{m,n}}}$ is the distance between object $m$ and $n$, and ${{L_e}}$ is the path-loss exponent.
In addition, the diagonal matrix ${{{\bf{F}}_C}}$ represents the path loss with ${\left[ {{{\bf{F}}_C}} \right]_{i,i}} = {\varepsilon _{A,i}}$ property. As antenna arrays of the BS and relay UAV are collocated, the diagonal matrix in ${{\bf{F}}_C}$ can be expressed as ${{\bf{F}}_M} = {\varepsilon _M}{{\bf{I}}_{{N_{UAV}}}}$.

Building on this, the channel state information estimation is performed at the BS and relay UAV, and the minimum mean square error (MMSE) estimates~\cite{timilsina2017secure} of ${\bf{M}}$ and ${\bf{N}}$ can be respectively calculated as
% \begin{equation}\label{162546}\small
% \begin{array}{l}
% {\bf{\hat M}} = \\
% {{\bf{F}}_M}{\left( {{{\bf{F}}_M} + {{{{\bf{I}}_{{N_{UAV}}}}} \mathord{\left/
%  {\vphantom {{{{\bf{I}}_{{N_{UAV}}}}} {\left( {{T_{UAV}}{P_{UAV}}} \right)}}} \right.
%  \kern-\nulldelimiterspace} {\left( {{T_{UAV}}{P_{UAV}}} \right)}}} \right)^{ - 1}}\left( {{\bf{M}} + {{{{\bf{A}}_M}} \mathord{\left/
%  {\vphantom {{{{\bf{A}}_M}} {\sqrt {{T_{UAV}}{P_{UAV}}} }}} \right.
%  \kern-\nulldelimiterspace} {\sqrt {{T_{UAV}}{P_{UAV}}} }}} \right)
% \end{array},
% \end{equation}
% \begin{align}\label{162546}
% {\bf{\hat M}} = \frac{{{{\bf{F}}_M}}}{{{{\bf{F}}_M} + {{{{\bf{I}}_{{N_{UAV}}}}} \mathord{\left/
%  {\vphantom {{{{\bf{I}}_{{N_{UAV}}}}} {{T_{UAV}}{P_{UAV}}}}} \right.
%  \kern-\nulldelimiterspace} {{T_{UAV}}{P_{UAV}}}}}} \times \left( {{\bf{M}} + \frac{{{{\bf{A}}_M}}}{{\sqrt {{T_{UAV}}{P_{UAV}}} }}} \right)
% \end{align}
\begin{align}\label{162546}
{\bf{\hat M}} = {{\bf{F}}_M}{\left( {{{\bf{F}}_M} + \frac{{{{\bf{I}}_{{N_{UAV}}}}}}{{{T_{UAV}}{P_{UAV}}}}} \right)^{ - 1}}\left( {{\bf{M}} + \frac{{{{\bf{A}}_M}}}{{\sqrt {{T_{UAV}}{P_{UAV}}} }}} \right)
\end{align}
and
\begin{equation}\label{562514895}
{{\bf{\hat N}} = {{\bf{F}}_N}{{\left( {{{\bf{F}}_N} + \frac{{{{\bf{I}}_{{N_{user}}}}}}{{{T_{user}}{P_{user}}}}} \right)}^{ - 1}} \times \left( {{\bf{N}} + \frac{{{{\bf{A}}_N}}}{{\sqrt {{T_{user}}{P_{user}}} }}} \right)},
\end{equation}
% \begin{equation}\label{562514895}
% {{\bf{\hat N}} = \frac{{{{\bf{F}}_N}}}{{\left( {{{\bf{F}}_N} + {{{{\bf{I}}_{{N_{user}}}}} \mathord{\left/
%  {\vphantom {{{{\bf{I}}_{{N_{user}}}}} {{T_{user}}{P_{user}}}}} \right.
%  \kern-\nulldelimiterspace} {{T_{user}}{P_{user}}}}} \right)}} \times \left( {{\bf{N}} + \frac{{{{\bf{A}}_N}}}{{\sqrt {{T_{user}}{P_{user}}} }}} \right)},
% \end{equation}
% \begin{equation}\label{562514895}\small
% \begin{array}{l}
% {\bf{\hat N}} = \\
% {{\bf{F}}_N}{\left( {{{\bf{F}}_N} + {{{{\bf{I}}_{{N_{user}}}}} \mathord{\left/
%  {\vphantom {{{{\bf{I}}_{{N_{user}}}}} {\left( {{T_{user}}{P_{user}}} \right)}}} \right.
%  \kern-\nulldelimiterspace} {\left( {{T_{user}}{P_{user}}} \right)}}} \right)^{ - 1}}\left( {{\bf{N}} + {{{{\bf{A}}_N}} \mathord{\left/
%  {\vphantom {{{{\bf{A}}_N}} {\sqrt {{T_{user}}{P_{user}}} }}} \right.
%  \kern-\nulldelimiterspace} {\sqrt {{T_{user}}{P_{user}}} }}} \right)
% \end{array},
% \end{equation}
where ${{\bf{A}}_M} \sim {\cal C}{{\cal N}_{{N_{UAV}} \times {N_{BS}}}}\left( {{{\bf{0}}_{{N_{UAV}} \times {N_{BS}}}},{{\bf{I}}_{{N_{UAV}}}} \otimes {{\bf{I}}_{{N_{BS}}}}} \right)$ and ${{\bf{A}}_N} \sim {\cal C}{{\cal N}_{{N_{user}} \times {N_{UAV}}}}\left( {{{\bf{0}}_{{N_{user}} \times {N_{UAV}}}},{{\bf{I}}_{{N_{user}}}} \otimes {{\bf{I}}_{{N_{UAV}}}}} \right)$. ${{T_{UAV}}}$ and ${{T_{user}}}$ represent the pilot transmission durations of the relay UAV and mobile users, respectively. ${{P_{UAV}}}$ and ${{P_{user}}}$ indicate the pilot transmission power at the relay UAV and mobile users, respectively. According to the orthogonality property of MMSE, ${\bf{N}}$ can be determined as:
\begin{equation}\label{4561456}
{\bf{N}} = {\bf{\tilde N}} + {{\bm{\xi }}_N},
\end{equation}
where ${{\bm{\xi }}_N}$ represents the estimation error matrix, and the rows of ${{\bf{\tilde N}}}$ and ${{\bm{\xi }}_N}$ possess the mutually independent and distributive property, indicating ${\bf{\tilde N}}\left( {j,:} \right) \sim {\cal C}{\cal N}\left( {{\bf{0}},{{{\bf{\tilde F}}}_N}} \right)$ and ${{\bf{\xi }}_N}\left( {j,:} \right) \sim {\cal C}{\cal N}\left( {{\bf{0}},{{\bf{F}}_N} - {{{\bf{\tilde F}}}_N}} \right)$. ${{{{\bf{\tilde F}}}_N}}$ is a diagonal matrix with the $k$-th element is calculated as:
\begin{equation}\label{7654646}
{{\tilde \varepsilon }_{{N_k}}} = {{\varepsilon _{{N_k}}^2} \mathord{\left/
 {\vphantom {{\varepsilon _{{N_k}}^2} {\left( {{\varepsilon _{{N_i}}} + {1 \mathord{\left/
 {\vphantom {1 {\left( {{T_{user}}{P_{user}}} \right)}}} \right.
 \kern-\nulldelimiterspace} {\left( {{T_{user}}{P_{user}}} \right)}}} \right)}}} \right.
 \kern-\nulldelimiterspace} {\left( {{\varepsilon _{{N_i}}} + {1 \mathord{\left/
 {\vphantom {1 {\left( {{T_{user}}{P_{user}}} \right)}}} \right.
 \kern-\nulldelimiterspace} {\left( {{T_{user}}{P_{user}}} \right)}}} \right)}}.
\end{equation}

Therefore, the zero force precoder in (\ref{11111}) can be further calculated as:
\begin{equation}\label{7455}
{{\bf{W}}_{\mathop{\rm c}\nolimits} } = \alpha {\left( {{\bf{\hat N\hat M}}} \right)^{\mathop{\rm H}\nolimits} }{\left( {{\bf{\hat N\hat M}}{{\left( {{\bf{\hat N\hat M}}} \right)}^{\mathop{\rm H}\nolimits} }} \right)^{ - 1}},
\end{equation}
where $\alpha$ denotes the power normalization factor, which is expressed as:
\begin{equation}\label{962452}\small
\alpha  = \frac{1}{{\sqrt {\mathbb{E}\left[ {{\mathop{\rm Tr}\nolimits} \left( {{{\left( {{\bf{\hat N\hat M}}{{\left( {{\bf{\hat N\hat M}}} \right)}^{\mathop{\rm H}\nolimits} }} \right)}^{ - 1}}} \right)} \right]} }}.
\end{equation}

On this basis, the received communication signal at the relay UAV is
\begin{equation}\label{48451}
{{\bf{y}}_{UAV}} = \sqrt {{P_{com}}} {\bf{M}}{{\bf{W}}_{\mathop{\rm c}\nolimits} }{{\bf{x}}_{\mathop{\rm c}\nolimits} } + \sqrt {{P_{AN}}} {\bf{M}}{{\bf{V}}_{\mathop{\rm n}\nolimits} }{{\bf{z}}_{\mathop{\rm n}\nolimits} } + {{\bf{n}}_{UAV}},
\end{equation}
where ${{{\bf{n}}_{UAV}}}$ is the additive white Gaussian noise (AWGN) vector, satisfying $\mathbb{E}\left[ {{{\bf{n}}_{UAV}}{{\bf{n}}_{UAV}}^{\mathop{\rm H}\nolimits} } \right] = \sigma _{UAV}^2{{\bf{I}}_{{N_{UAV}}}}$. Accordingly, the amplification factor of the relay UAV can be calculated as
\begin{equation}\label{48566}\small
\begin{array}{l}
\beta  = \\
\sqrt {\frac{{{P_{UAV}}}}{{{P_{com}}\mathbb{E}\left[ {{\mathop{\rm Tr}\nolimits} \left( {{\bf{M}}{{{\bf{\hat W}}}_{\mathop{\rm c}\nolimits} }{{\left( {{\bf{M}}{{{\bf{\hat W}}}_{\mathop{\rm c}\nolimits} }} \right)}^{\mathop{\rm H}\nolimits} }} \right)} \right] + {P_{AN}}\sigma _Z^2\mathbb{E}\left[ {{\mathop{\rm Tr}\nolimits} \left( {{\bf{M}}{{\bf{V}}_{\mathop{\rm n}\nolimits} }{{\left( {{\bf{M}}{{\bf{V}}_{\mathop{\rm n}\nolimits} }} \right)}^{\mathop{\rm H}\nolimits} }} \right)} \right] + {N_{UAV}}\sigma _{UAV}^2}}}
\end{array},
\end{equation}
and thereby the signal transmitted by the relay UAV is
\begin{equation}\label{1564454}
\begin{array}{l}
{{\bf{x}}_{UAV}} = \beta {{\bf{y}}_{UAV}}\\
 = \beta \left( {\sqrt {{P_{com}}} {\bf{M}}{{\bf{W}}_{\mathop{\rm c}\nolimits} }{{\bf{x}}_{\mathop{\rm c}\nolimits} } + \sqrt {{P_{AN}}} {\bf{M}}{{\bf{V}}_{\mathop{\rm n}\nolimits} }{{\bf{z}}_{\mathop{\rm n}\nolimits} } + {{\bf{n}}_{UAV}}} \right)
\end{array},
\end{equation}

After that, the received signal at the mobile user and the $l$-th mobile user can be expressed as
\begin{align}\label{5641233485646}
{{\bf{y}}_{user}} &= \beta \sqrt {{P_{com}}} {\bf{NM}}{{{\bf{\hat W}}}_{\mathop{\rm c}\nolimits} }{{\bf{x}}_{\mathop{\rm c}\nolimits} } + \beta \sqrt {{P_{AN}}} {\bf{NM}}{{\bf{V}}_{\mathop{\rm n}\nolimits} }{{\bf{z}}_{\mathop{\rm n}\nolimits} }\\ \notag
 &+ \beta {\bf{N}}{{\bf{n}}_{UAV}} + {{\bf{n}}_{user}}
\end{align}
and
\begin{align}\label{26235113231224242}
{{\bf{y}}_{user}}_{,l} &=\beta \sqrt {{P_{com}}} {{\bf{n}}_l}{\bf{M}}{{{\bf{\hat w}}}_{{\mathop{\rm c}\nolimits} ,l}}{{\bf{x}}_{{\mathop{\rm c}\nolimits} ,l}} \\ \notag
&+ \sum\nolimits_{j = 1,j \ne l}^{{N_{user}}} {\beta \sqrt {{P_{com}}} {{\bf{n}}_j}{\bf{M}}{{{\bf{\hat w}}}_{{\mathop{\rm c}\nolimits} ,j}}{{\bf{x}}_{{\mathop{\rm c}\nolimits} ,j}}} \\ \notag
 &+ \beta \sqrt {{P_{AN}}} {{\bf{n}}_l}{\bf{M}}{{\bf{V}}_{\mathop{\rm n}\nolimits} }{{\bf{z}}_{\mathop{\rm n}\nolimits} } + \beta {{\bf{n}}_l}{{\bf{n}}_{UAV}} + {{\bf{n}}_{user,l}},
\end{align}
respectively, where ${{{\bf{n}}_{user}}}$ represents the AWGN vector that satisfies $\mathbb{E}\left[ {{{\bf{n}}_{user}}{{\bf{n}}_{user}}^{\mathop{\rm H}\nolimits} } \right] = \sigma _{user}^2{{\bf{I}}_{{N_{user}}}}$.

In this case, the eavesdropper can intercept the transmitted signal from the BS. Hence, the received signals and the intercepted signal of the $j$-th mobile user can be respectively represented as
\begin{equation}\label{56446485646}
{{\bf{y}}_{eave}}_1 = \sqrt {{P_{com}}} {\bf{Q}}{{{\bf{\hat W}}}_{\mathop{\rm c}\nolimits} }{{\bf{x}}_{\mathop{\rm c}\nolimits} } + \sqrt {{P_{AN}}} {\bf{Q}}{{\bf{V}}_{\mathop{\rm n}\nolimits} }{{\bf{z}}_{\mathop{\rm n}\nolimits} } + {{\bf{n}}_{eave}},
\end{equation}
and 
\begin{align}\label{26235124242}
{{\bf{y}}_{eave}}_{1,l} &= \sqrt {{P_{com}}} {{\bf{q}}_l}{{{\bf{\hat w}}}_{\mathop{\rm c}\nolimits} }{{\bf{x}}_{\mathop{\rm c}\nolimits} }\\ \notag
&+ \sum\nolimits_{j = 1,j \ne l}^{{N_{user}}} {\beta \sqrt {{P_{com}}} {{\bf{q}}_j}{{{\bf{\hat w}}}_{{\mathop{\rm c}\nolimits} ,j}}{{\bf{x}}_{{\mathop{\rm c}\nolimits} ,j}}} \\ \notag
 &+ \sqrt {{P_{AN}}} {{\bf{q}}_l}{{\bf{V}}_{\mathop{\rm n}\nolimits} }{{\bf{z}}_{\mathop{\rm n}\nolimits} } + {{\bf{n}}_{eave,l}},
\end{align}
where ${{{\bf{n}}_{eave}}}$ represents the AWGN vector at the eavesdropper that satisfies
$\mathbb{E}\left[ {{{\bf{n}}_{eave}}{{\bf{n}}_{eave}}^{\mathop{\rm H}\nolimits} } \right] = \sigma _{eave}^2{{\bf{I}}_{{N_{eave}}}}$.

In the same way, the eavesdropper can also intercept the transmitted information from the relay UAV. The received signal and the intercepted $l$-th mobile user's signal can be respectively expressed as:
\begin{align}\label{963255}
{{\bf{y}}_{eave}}_2 &= \beta \sqrt {{P_{com}}} {\bf{ZM}}{{{\bf{\hat W}}}_{\mathop{\rm c}\nolimits} }{{\bf{x}}_{\mathop{\rm c}\nolimits} } + \beta \sqrt {{P_{AN}}} {\bf{ZM}}{{\bf{V}}_{\mathop{\rm n}\nolimits} }{{\bf{z}}_{\mathop{\rm n}\nolimits} }\\ \notag
 &+ \beta {\bf{Z}}{{\bf{n}}_{UAV}} + {{\bf{n}}_{eave}},
\end{align}
and
\begin{align}\label{96214785}
{{\bf{y}}_{eave}}_{2,l}&=\beta \sqrt {{P_{com}}} {{\bf{z}}_l}{\bf{M}}{{{\bf{\hat w}}}_{{\mathop{\rm c}\nolimits} ,l}}{{\bf{x}}_{{\mathop{\rm c}\nolimits} ,l}}\\ \notag
&+ \sum\nolimits_{j = 1,j \ne l}^{{N_{user}}} {\beta \sqrt {{P_{com}}} {{\bf{z}}_j}{\bf{M}}{{{\bf{\hat w}}}_{{\mathop{\rm c}\nolimits} ,j}}{{\bf{x}}_{{\mathop{\rm c}\nolimits} ,j}}} \\ \notag
 &+ \beta \sqrt {{P_{AN}}} {{\bf{z}}_l}{\bf{M}}{{\bf{V}}_{\mathop{\rm n}\nolimits} }{{\bf{z}}_{\mathop{\rm n}\nolimits} } + \beta {{\bf{z}}_l}{{\bf{n}}_{UAV}} + {{\bf{n}}_{eave,l}}.
\end{align}

For the $l$-th mobile user, the received signal includes the desired signal and uncorrelated noise, which is expressed as
\begin{equation}\label{5615315}
{y_{user,l}} = \beta \sqrt {{P_{com}}} \mathbb{E}\left[ {{{\bf{n}}_l}{\bf{M}}{{{\bf{\hat w}}}_{{\mathop{\rm c}\nolimits} ,l}}} \right]{{\bf{x}}_{{\mathop{\rm c}\nolimits} ,l}} + {\vartheta _l},
\end{equation}
where ${\vartheta _l}$ means the effective noise signal, calculated as:
\begin{align}\label{486456145}
{\vartheta _l} &= \beta \sqrt {{P_{com}}} \left( {{{\bf{n}}_l}{\bf{M}}{{{\bf{\hat w}}}_{{\mathop{\rm c}\nolimits} ,l}} - \mathbb{E}\left[ {{{\bf{n}}_l}{\bf{M}}{{{\bf{\hat w}}}_{{\mathop{\rm c}\nolimits} ,l}}} \right]} \right){{\bf{x}}_{{\mathop{\rm c}\nolimits} ,l}}\\ \notag
 &+ \sum\nolimits_{j = 1,j \ne l}^{{N_{user}}} {\beta \sqrt {{P_{com}}} {{\bf{n}}_l}{\bf{M}}{{{\bf{\hat w}}}_{{\mathop{\rm c}\nolimits} ,j}}{{\bf{x}}_{{\mathop{\rm c}\nolimits} ,j}}} \\ \notag
 &+ \beta \sqrt {{P_{com}}} {{\bf{n}}_l}{\bf{M}}{{\bf{V}}_{\mathop{\rm n}\nolimits} }{{\bf{z}}_{\mathop{\rm n}\nolimits} }\\ \notag
 &+ \beta {{\bf{n}}_l}{{\bf{n}}_{UAV}} + {{\bf{n}}_{user,l}},
\end{align}

Following \cite{zhu2015linear}, to mitigate the risk of information interception by the eavesdropper, a null-space-based AN precoder is employed, and the BS’s AN shaping matrix is calculated as
\begin{equation}\label{256651456}
{{\bf{V}}_{\mathop{\rm n}\nolimits} } = \frac{{\bf{D}}}{{\sqrt {\mathbb{E}\left[ {{\mathop{\rm Tr}\nolimits} \left[ {{\bf{D}}{{\bf{D}}^{\mathop{\rm H}\nolimits} }} \right]} \right]} }},
\end{equation}
where ${\bf{D}}$ is 
\begin{equation}\label{456523212}
{\bf{D}} = {{\bf{I}}_{{N_{UAV}}}} - {\left( {{\bf{\hat N\hat M}}} \right)^{\mathop{\rm H}\nolimits} }{\left( {{\bf{\hat N\hat M}}{{\left( {{\bf{\hat N\hat M}}} \right)}^{\mathop{\rm H}\nolimits} }} \right)^{ - 1}}{\bf{\hat N\hat M}}.
\end{equation}

Based on the above analysis, the achievable rate at the $l$-th mobile user can be defined as
\begin{align}\label{7798531212}
{\gamma _{user,l}} &= \frac{{\min \left( {{T_{c,UAV}} - {T_{UAV}},{T_{c,BS}} - {T_{BS}}} \right)}}{{{T_{c,UAV}} + {T_{c,BS}}}} \\ \notag
& \times \log \left( {1 + \frac{{{\beta ^2}{P_{com}}{{\mathbb{E}\left| {\left[ {{{\bf{z}}_i}{\bf{M}}{{{\bf{\hat w}}}_{{\mathop{\rm c}\nolimits} ,l}}} \right]} \right|}^2}}}{{\sum\nolimits_{j = 1}^4 {{P_{{\mathop{\rm interf}\nolimits} ,j}} + \sigma _{user}^2} }}} \right),
\end{align}
where ${{T_{c,UAV}}}$ and ${{T_{c,BS}}}$ represent the coherence time intervals from relay UAV to BS, and mobile users to relay UAV, respectively. Besides, ${P_{{\mathop{\rm interf}\nolimits} ,j}}$ can be calculated as
\begin{equation}\label{561251511}
{P_{{\mathop{\rm interf}\nolimits} ,1}} = {\beta ^2}{P_{com}}{\mathop{\rm Var}\nolimits} \left[ {{{\bf{z}}_i}{\bf{M}}{{{\bf{\hat w}}}_{{\mathop{\rm c}\nolimits} ,i}}} \right],
\end{equation}
\begin{equation}\label{561251522}
{P_{{\mathop{\rm interf}\nolimits} ,2}} = \sum\nolimits_{j = 1,j \ne i}^{{N_{user}}} {{\beta ^2}{P_{com}}} \mathbb{E}\left[ {{{\left\| {{{\bf{z}}_i}{\bf{M}}{{{\bf{\hat w}}}_{{\mathop{\rm c}\nolimits} ,i}}} \right\|}^2}} \right],
\end{equation}
\begin{equation}\label{561251533}
{P_{{\mathop{\rm interf}\nolimits} ,3}} = {\beta ^2}\sigma _Z^2{P_{com}}\mathbb{E}\left[ {{{\left\| {{{\bf{z}}_i}{\bf{M}}{{\bf{V}}_n}} \right\|}^2}} \right],
\end{equation}
\begin{equation}\label{561251544}
{P_{{\mathop{\rm interf}\nolimits} ,4}} = {\beta ^2}\mathbb{E}\left[ {{{\left\| {{{\bf{z}}_i}{{\bf{n}}_{UAV}}} \right\|}^2}} \right],
\end{equation}
where~(\ref{561251511}) indicates the interference due to imperfect channel estimation,~(\ref{561251522}) is the inter-user interference (IUI) from other served users,~(\ref{561251533}) is caused by the artificial noise, and~(\ref{561251544}) is the amplified noise from the relay.

The data transmission process from BS to mobile users can be divided into two timeslots: the first one is BS to relay UAV and the second one from relay UAV to mobile users. For these timeslots, the rate leaked to the eavesdropper are determined as
\begin{equation}\label{26561455}
\begin{array}{l}
{\gamma _{eave1,l}} = \\
\log \left( {1 + \frac{{{P_{com}}\mathbb{E}\left[ {{{\left\| {{{\bf{q}}_l}{{{\bf{\hat w}}}_{{\mathop{\rm c}\nolimits} ,l}}} \right\|}^2}} \right]}}{{\sum\nolimits_{j = 1,j \ne l}^{{N_{user}}} {{P_{com}}\mathbb{E}\left[ {{{\left\| {{{\bf{q}}_i}{{{\bf{\hat w}}}_{{\mathop{\rm c}\nolimits} ,l}}} \right\|}^2}} \right] + \sigma _Z^2{P_{AN}}\mathbb{E}\left[ {{{\left\| {{{\bf{z}}_l}{{\bf{V}}_n}} \right\|}^2}} \right] + \sigma _{eave}^2} }}} \right)
\end{array},
\end{equation}
and
\begin{equation}\label{154154}
{\gamma _{eave2,l}} = \log \left( {1 + \frac{{{\beta ^2}{P_{com}}\mathbb{E}\left[ {{{\left| {{{\bf{z}}_l}{\bf{M}}{{{\bf{\hat w}}}_{{\mathop{\rm c}\nolimits} ,l}}} \right|}^2}} \right]}}{{\sum\nolimits_{j = 1}^3 {{P_{J,j}} + \sigma _{eave}^2} }}} \right),
\end{equation}
respectively, where for ${{P_{J,j}}}$ we have
\begin{equation}\label{156151511}
{P_{J,1}} = \sum\nolimits_{j = 1,j \ne l}^{{N_{user}}} {{\beta ^2}{P_{com}}} \mathbb{E}\left[ {{{\left\| {{{\bf{z}}_l}{\bf{M}}{{{\bf{\hat w}}}_{{\mathop{\rm c}\nolimits} ,l}}} \right\|}^2}} \right],
\end{equation}
\begin{equation}\label{156151522}
{P_{J,2}} = {\beta ^2}\sigma _Z^2{P_{com}}\mathbb{E}\left[ {{{\left\| {{{\bf{z}}_l}{\bf{M}}{{\bf{V}}_n}} \right\|}^2}} \right],
\end{equation}
\begin{equation}\label{156151533}
{P_{J,3}} = {\beta ^2}\mathbb{E}\left[ {{{\left\| {{{\bf{z}}_l}{{\bf{n}}_{UAV}}} \right\|}^2}} \right].
\end{equation}

Based on (\ref{26561455}) and (\ref{154154}), the cumulative information leakage rate to the eavesdropper is calculated as
\begin{equation}\label{15475452435}
{\gamma _{eave,l}} = \frac{1}{2}\left( {{\gamma _{eave1,l}} + {\gamma _{eave2,l}}} \right).
\end{equation}
Therefore, the secrecy rate of the secrecy communications can be obtained
\begin{equation}\label{1415456123123}
{\gamma _{secure,l}} = \max \left[ {0,\left( {{\gamma _{user,l}} - {\gamma _{eave,l}}} \right)} \right].
\end{equation}

\subsection{Data Freshness Modeling for Computation}

\begin{figure}[!htb]
  \centering
  \includegraphics[width=8.7cm]{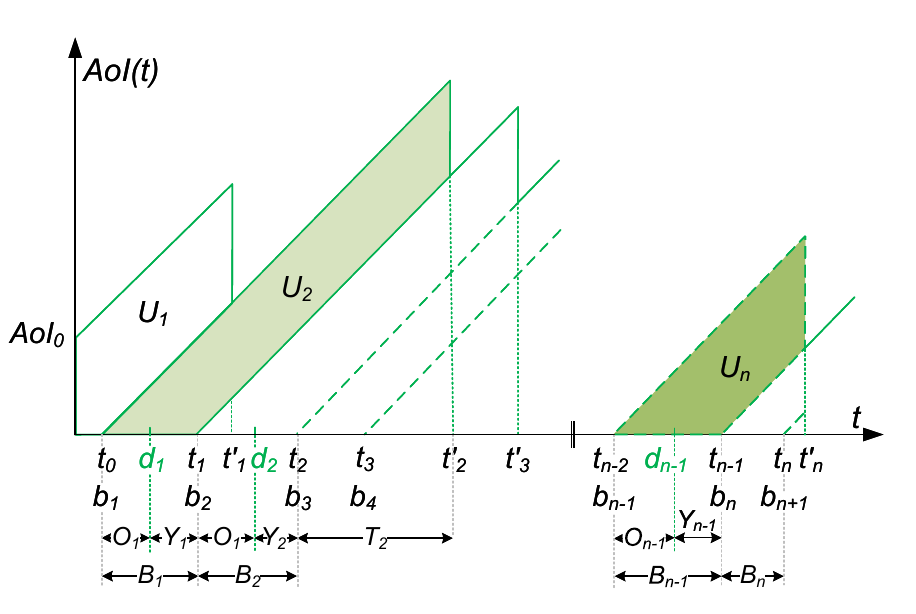}\\
  \caption{The computation-oriented average AoI model.}
  \label{AoImodel2}
\end{figure}

% To quantitatively analyze the data freshness of the communication and computation process, average AoI performance metric is adopted. Specifically, as shown in Fig. \ref{AoImodel2}, according to [xxx], following the zero-wait policy, when the computing time is exponentially distributed and under the first-come-first-serve (FCFS) serving rule, $b_i$ means the generation time of the $i$-th data, also representing the time instant $t_{i-1}$ when the $(i-1)$-th data arrives at the relay UAV computing queue. $d_i$ is denoted as the time instant when the $i$-th data is computed by the BS, and ${{t'}_i}$ represents the computing time instant at the relay UAV end. In this regard, when AoI is the sawtooth function, it can be mathematically expressed as:
As shown in Fig. \ref{AoImodel2} and following, under the zero-wait policy with exponentially distributed computing times and a first-come-first-serve (FCFS) rule \cite{kuang2020analysis}, $b_i$ denotes the generation time of the $i$-th data packet, which also corresponds to the arrival instant $t_{i-1}$ of the $(i-1)$-th packet at the relay UAV’s computing queue. $d_i$ represents the time when the $i$-th packet is computed by the BS, and $t'_i$ indicates the computing time at the relay UAV. In this case, the AoI can be mathematically expressed as $AoI\left( t \right) = t - \Delta \left( t \right)$, where $\Delta \left( t \right)$ is the time-stamp of the data generation. At the initial time stamp, the AoI is given by $AoI\left( 0 \right) = Ao{I_0}$. On this basis, the average AoI is calculated as the unit area surrounded by the AoI function and the time axis \cite{yang2023age}. Mathematically, within the time period $\left[ {0,\tau } \right]$, the average AoI is expressed as:
% \begin{equation}\label{261455445}
% AoI\left( t \right) = t - \Delta \left( t \right),
% \end{equation}

\begin{equation}\label{26564564}
AAo{I_\tau } = \frac{1}{\tau }\int_0^\tau  {AoI\left( t \right)dt} .
\end{equation}

Let the observation interval be $\tau  = {{t'}_n}$, then the average AoI can be written as:
\begin{equation}\label{3514313}
AAo{I_\tau } = \frac{1}{\tau }\left[ {\sum\nolimits_{i = 1}^n {{U_i} + \frac{1}{2}{{\left( {{B_n} + {T_n}} \right)}^2}} } \right].
\end{equation}

% As can be seen from Fig. \ref{AoImodel2}, $U_1$ is a polygon, and for ${U_i}\left( {i \ge 2} \right)$ belongs to the isosceles trapezoid, which is calculated as:
As shown in Fig. \ref{AoImodel2}, $U_1$ is a polygon, while for $i \ge 2$, ${U_i}$ takes the form of an isosceles trapezoid. Hence, we have 
\begin{align}\label{78451512312}
{U_i} &= \frac{1}{2}{\left( {{B_{i - 1}} + {B_i} + {T_i}} \right)^2} - \frac{1}{2}{\left( {{B_i} + {T_i}} \right)^2}\\ \notag
 &= \frac{1}{2}B_{i - 1}^2 + {T_i}{B_{i - 1}} + {B_{i - 1}}{B_i},
\end{align}
where ${B_i} = {b_{i + 1}} - {b_i} = {t_i} - {t_{i - 1}}$ is the inter-generation time between the $i$-th and $(i-1)$-th data packet at the BS. ${B_i}$ also represents the data' computing time at the BS and its transmission time from the BS to the receiver, indicating ${B_i} = {O_i} + {Y_i}$. Here, ${O_i} = {d_i} - {b_i}$ denotes the service time of the $i$-th data packet at the BS and ${Y_i} = {t_i} - {d_i}$ is the service time of wireless transmission channel. Let ${T_i} = {{t'}_i} - {t_i}$ represent the elapsed time for the $i$-th status-updating data packet, measured from its arrival at the relay UAV's computing queue to the completion of its service. Then, the average AoI can be further calculated as
\begin{equation}\label{4863123123}
AAo{I_\tau } = \frac{{\tilde U}}{\tau } + \frac{{n - 1}}{\tau }\frac{1}{{n - 1}}\sum\nolimits_{i = 2}^n {{U_i}},
\end{equation}
where $\tilde U = {U_1} + \frac{1}{2}{\left( {{B_n} + {T_n}} \right)^2}$. When $\tau  \to \infty $, the first term in (\ref{4863123123}) can be negligible. Besides, Fig.~\ref{AoImodel2} shows ${{t'}_n} = {t_0} + \sum\nolimits_{i = 1}^n {\left( {{O_i} + {Y_i}} \right)}  + {T_n}$, we have $\mathop {\lim }\limits_{\tau  \to \infty } \frac{\tau }{n} = \mathbb{E}\left[ {{O_i} + {Y_i}} \right]$. ${\tau  \mathord{\left/
 {\vphantom {\tau  {\left( {n - 1} \right)}}} \right.
 \kern-\nulldelimiterspace} {\left( {n - 1} \right)}}$ can be regarded as sum service time of BS's computing and transmission process, and then
\begin{equation}\label{1512123}
\mathop {\lim }\limits_{\tau  \to \infty } \frac{\tau }{n} = \frac{1}{{{\mu _{BS}}}} + \frac{1}{{{\mu _{trans}}}},
\end{equation}
where $\mu _{trans}$ is the transmission time\footnote{Note that as the computed data size is small enough, based on \cite{han2021age}, the transmission time period from the relay UAV to the mobile users can be negligible. Besides, we treat the secrecy rate of Eq.  \ref{1415456123123} as the wireless channel's service rate within the whole data computing process.}, and ${\mu _{trans}} = {\gamma _{secure,l}} > 0$, and ${{\mu _{BS}}}$ is the data computing rate of the BS. On this basis, the average AoI can be determined as:
\begin{equation}\label{21212}
\begin{array}{l}
AAoI = \mathop {\lim }\limits_{\tau  \to \infty } Ao{I_\tau } = \frac{{{\mu _{BS}}{\mu _{trans}}}}{{{\mu _{BS}} + {\mu _{trans}}}}\mathbb{E}\left[ {{U_i}} \right]\\
 = \frac{{{\mu _{BS}}{\mu _{trans}}}}{{{\mu _{BS}} + {\mu _{trans}}}}\left( {\frac{1}{2}\mathbb{E}\left[ {B_{i - 1}^2} \right] + \mathbb{E}\left[ {{T_i}{B_{i - 1}}} \right] + \mathbb{E}\left[ {{B_{i - 1}}{B_i}} \right]} \right)
\end{array}.
\end{equation}

\textbf{Theorem 1:} When both the BS and the relay UAV's data computing are exponentially distributed, then, under the FCFS rule and zero-wait policy, the average AoI of the data can be calculated as
\begin{equation}\label{456123145456452}
\begin{array}{l}
AAoI = \frac{1}{{{\mu _{UAV}}}} + \frac{{{\varpi _1} - 1}}{{{\mu _{BS}}}} - \frac{{1 + {\varpi _2}}}{{{\mu _{trans}}}} + \frac{{{\mu _{BS}}{\mu _{trans}}}}{{{\mu _{BS}} + {\mu _{trans}}}}\\
\left( {\frac{1}{{{\mu _{BS}}{\mu _{UAV}}}}} \right. + \frac{1}{{{\mu _{trans}}{\mu _{UAV}}}} + \frac{2}{{\mu _{BS}^2}} + \frac{2}{{\mu _{trans}^2}} + \frac{3}{{{\mu _{BS}}{\mu _{trans}}}}\\
 + \frac{{{\mu _{BS}}{\mu _{trans}}}}{{{\mu _{trans}} - {\mu _{BS}}}}\left( {\frac{1}{{{{\left( {{\varpi _3} + {\mu _{BS}}} \right)}^2}}} - \frac{1}{{{{\left( {{\varpi _3} + {\mu _{trans}}} \right)}^2}}}} \right)\\
\left. {\left( {\frac{1}{{{\varpi _3}}} - \frac{{{\varpi _1}}}{{{\varpi _3} + {\mu _{BS}}}} + \frac{{{\varpi _2}}}{{{\varpi _3} + {\mu _{trans}}}}} \right)} \right)
\end{array},
\end{equation}
where ${\varpi _1}$, ${\varpi _2}$, and ${\varpi _3}$ are defined as
\begin{equation}\label{151514512}
{\varpi _1} = \frac{{{\mu _{UAV}}{\mu _{trans}}}}{{\left( {{\mu _{trans}} - {\mu _{BS}}} \right)\left( {{\mu _{BS}} + {\mu _{UAV}}} \right)}},
\end{equation}
\begin{equation}\label{151514523}
{\varpi _2} = \frac{{{\mu _{BS}}{\mu _{trans}}}}{{\left( {{\mu _{trans}} - {\mu _{BS}}} \right)\left( {{\mu _{trans}} + {\mu _{UAV}}} \right)}},
\end{equation}
and 
\begin{equation}\label{151514534}
\begin{array}{l}
{\varpi _3} = \frac{1}{2}\left[ {{\mu _{UAV}} - \left( {{\mu _{BS}} + {\mu _{trans}}} \right)} \right.\\
\left. { + \sqrt {{{\left( {{\mu _{UAV}} - {\mu _{trans}} + {\mu _{BS}}} \right)}^2} + 4{\mu _{trans}}{\mu _{UAV}}} } \right]
\end{array},
\end{equation}
respectively.
\begin{IEEEproof}
Please refer to Appendix A.
\end{IEEEproof}

\subsection{ISCC performances optimization problem formulation}
To ensure ISCC performances of LAWNs, metrics including sensing beampattern error, secret communication rate, AoI for computation, and UAV's computing energy need to be optimized simultaneously. As BS's transmitting power allocation and computing strategies directly influence the above mentioned performance metrics, we aim at jointly optimizing $P_{BS}$, $P_{sens}$, $\mu_{BS}$ and $\mu_{UAV}$, and formulate the following optimization problem
\begin{align}\small
\mathop {\min }\limits_{\scriptstyle{P_{BS}},{P_{radar}}\hfill\atop
\scriptstyle{\mu _{BS}},{\mu _{UAV}}\hfill} R{B_{error}} - {\varsigma _1}{\gamma _{secure}} + {\varsigma _2}AAoI + {\varsigma _3}\mu _{UAV}^3 \label{4874515}\\
s.t.\quad
{P_{BS}} + {P_{radar}} + {P_{AN}} = {P_{sum}} ,\tag{\ref{4874515}{a}} \label{4874515a}\\
0 < {P_{com}} < {P_{sum}} , \tag{\ref{4874515}{b}} \label{4874515b}\\
0 < {P_{sens}} < {P_{sum}} , \tag{\ref{4874515}{c}} \label{4874515c}\\
0 < \frac{{{\mu _{trans}}\left( {{\mu _{BS}} - {\mu _{UAV}}} \right)}}{{{\mu _{BS}}{\mu _{UAV}}}} < 1, \tag{\ref{4874515}{d}} \label{4874515d}\\
{\mu _{trans}} > {\gamma _{th}} \tag{\ref{4874515}{e}} \label{4874515e},
\end{align}

where ${\varsigma _1}$, ${\varsigma _2}$ and ${\varsigma _3}$ are the coefficients to balance units of different metrics, ${\varsigma _3}\mu _{UAV}^3$ indicates the computing energy consumption of UAV~\cite{jiang2025optimization}. As can be seen, constraints from (\ref{4874515a}) to (\ref{4874515c}) are for the transmission power of the BS. The constraint (\ref{4874515d}) is to ensure the steady state of the queue system, where $\rho  = {{{\mu _{trans}}\left( {{\mu _{BS}} - {\mu _{UAV}}} \right)} \mathord{\left/
 {\vphantom {{{\mu _{trans}}\left( {{\mu _{BS}} - {\mu _{UAV}}} \right)} {{\mu _{BS}}{\mu _{UAV}}}}} \right.
 \kern-\nulldelimiterspace} {{\mu _{BS}}{\mu _{UAV}}}}$ is defined as the utilization rate of the server within the whole queue system. The last constraint (\ref{4874515e}) guarantees the quality of service for the mobile user, requiring the communication transmission rate to exceed the threshold $\gamma_{th}$.

% \section{Deep reinforcement learning-based multi-objective evolutionary algorithm for ISCC performances optimization}
\section{DRL-Based Multi-Objective Evolutionary Optimization for ISCC}

\subsection{Basic Principle}

The optimization problem formulated in (\ref{4874515}) involves multiple conflicting objectives and intricate constraints. For instance, allocating more transmit power to sensing reduces the beampattern error but simultaneously decreases the secrecy communication rate and increases the average AoI. Similarly, raising the UAV’s computing rate can decrease average AoI, yet this comes at the cost of higher energy consumption at the UAV. Therefore, (\ref{4874515}) is essentially a constrained multi-objective optimization problem (CMOOP), for which multiobjective evolutionary algorithms (MOEAs) provide effective solution methods. Compared with traditional convex optimization~\cite{zhang2017securing} and DRL approaches~\cite{khurshid2023drl}, MOEAs offer lower computational complexity, strong adaptability, and ease of implementation. They explore the Pareto optimal solution set, furnish decision makers with a portfolio of alternatives, and enable explicit trade-offs among competing objectives. Moreover, MOEAs do not require gradient information, making them suitable when the objective functions are complex, non-differentiable, or unknown.

While effective, the utilization of the fixed evolution operator limits MOEA's performance. Hence, this paper proposes the deep Q-network (DQN)-based MOEA, which can automatically manage online evolutionary operator selection in response to the optimization problem's evolving characteristics, to achieve better ISCC performance.

\subsection{DQN-based MOEA for ISCC performance optimization}

\subsubsection{Preliminaries of DQN-based MOEA} For the CMOOP, it can be uniformly represented as:
\begin{equation}\label{151521231}
\begin{array}{l}
\min \quad {\bf{F}}\left( {\bf{X}} \right) = \left( {{f_1}\left( {\bf{X}} \right),{f_2}\left( {\bf{X}} \right), \ldots ,{f_m}\left( {\bf{X}} \right)} \right)\\
\;\quad \quad s.t.\;\;\;{g_i}\left( {\bf{X}} \right) \le 0,\quad i = 1,2, \ldots p\\
\quad \quad \quad \quad {h_j}\left( {\bf{X}} \right) = 0,\quad j = 1,2, \ldots q\\
\quad \quad \quad \quad {\bf{X}} \in \mathbb{R}
\end{array},
\end{equation}
where $m$ is the number of optimization objectives, ${\bf{X}} = {\left( {{x_1},{x_2}, \ldots ,{x_n}} \right)^T}$ represents $n$ decision variables, ${g_i}\left( {\bf{X}} \right)$ and ${h_i}\left( {\bf{X}} \right)$ mean the inequality and equality constraints with $p$ and $q$ are the number of corresponding constraints, respectively. During the training process, the loss function can be defined as
\begin{equation}\label{2315231231}
Loss = \frac{1}{{\left| {TD} \right|}}{\sum\nolimits_{td \in TD} {\left( {Q\left( {{s_{td}},{a_{td}}} \right) - {q_{td}}} \right)} ^2},
\end{equation}
where ${TD}$ represents the training data randomly sampled from the experience replay, ${\left( {{s_{td}},{a_{td}}} \right)}$ and ${Q\left( {{s_{td}},{a_{td}}} \right)}$ are the input and output of the deep Q-network, respectively. Besides, ${{q_{td}}}$ is the Q-value of taking action ${{a_{td}}}$ at state ${{s_{td}}}$, which can be expressed as
\begin{equation}\label{6735}
{q_{td}} = {r_{td}} + \xi \mathop {\max }\limits_{a' \in A} Q\left( {{s_{td + 1}},{{a'}_{td}}} \right),
\end{equation}
where $\xi $ is the hyper-parameter, $Q\left( {{s_{td + 1}},{{a'}_{td}}} \right)$ is the obtained maximum reward after executing action ${{a_{td}}}$.

\begin{figure}[t]
  \centering
  \includegraphics[width=8cm]{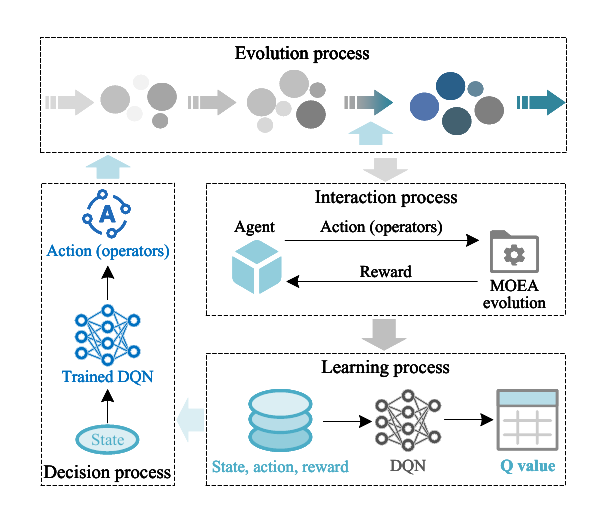}\\
  \caption{Diagram of the proposed DRL-based MOEA. The Evolution process uses evolutionary operators chosen by the DQN agent to modify the population. The interaction process shows how the agent controls MOEA evolution and receives rewards that reflect the impact of its actions. The Learning process then uses the resulting state-action-reward data to update the Q-values, improving its decision-making over time. The Decision process uses the current state of the population to inform the DQN's action selection. }
  \label{DRL4MOEA}
\end{figure}

Based on above preliminaries, as illustrated in Fig.~\ref{DRL4MOEA}, we integrate DQN into MOEA to enhance the operator selection when solving the CMOOP. Specifically, a DQN agent interacts with the MOEA environment. The state presented to the DQN agent is a vector consisting of convergence, diversity, and feasibility metrics, characterizing the population's performance in the current generation \footnote{These metrics are chosen as the state representation because they collectively provide a comprehensive and informative snapshot of the evolutionary search process. Each metric addresses a critical aspect of population-based optimization, allowing the DQN agent to make well-informed decisions about which evolutionary operator to apply.}. Based on this state, the DQN agent selects an action, which corresponds to choosing a specific evolutionary operator from a pre-defined set $\mathcal{A}$. The agent then receives a reward calculated as the improvement in the population's convergence, diversity, and feasibility performances. This reward guides the DQN to learn an optimal policy that maximizes both the convergence and diversity of the population while minimizing constraint violations. Finally, the agent stores the tuple in its experience replay buffer, and periodically samples mini-batches from this buffer to update the weights of its Q-network using the Bellman equation, improving its Q-value estimations and refining the operator selection policy. On this basis, the action set ${\cal A}$ is defined as follows:
\begin{equation}\label{5145}
{\cal A} = \left\{ {o{p_1},o{p_2}, \ldots ,o{p_{\left| {\cal A} \right|}}} \right\},
\end{equation}
where ${o{p_u}}$ means the $u$-th operator, and ${{\left| {\cal A} \right|}}$ represents the number of operators. In addition, the state ${\cal S}$ depends on the convergence $con$, diversity $div$, and feasibility $fea$ performances of the population \cite{ming2024constrained}, which are 
\begin{equation}\label{51515}
con = \frac{1}{\chi }\sum\nolimits_{{\bf{X}} \in POP} {\sum\nolimits_{i = 1}^m {{f_i}\left( {\bf{X}} \right)} } ,
\end{equation}
\begin{equation}\label{564894}
fea = \frac{1}{\chi }\sum\nolimits_{{\bf{X}} \in POP} {\phi \left( {\bf{X}} \right)}  = \frac{1}{\chi }\sum\nolimits_{{\bf{X}} \in POP} {\sum\nolimits_{j = 1}^q {{\varphi _j}\left( {\bf{X}} \right)} } ,
\end{equation}
\begin{equation}\label{5621554145}
div = \frac{1}{{\sum\nolimits_{i = 1}^m {\left( {f_j^{\max } - f_j^{\min }} \right)} }},
\end{equation}
respectively, where $\chi $ is the normalization factor, ${f_j^{\max }}$ and ${f_j^{\min }}$ represent the maximum and minimum objective function value of the $j$-th objective function, respectively. When $\iota$ denotes a small enough positive number, ${\varphi _v}\left( {\bf{X}} \right)$ is the degree of the $v$-th constraint violation of a solution $\bf{X}$, ${\varphi _v}\left( {\bf{X}} \right)$ can be calculated as
\begin{equation}\label{25732572}
{\varphi _v}\left( {\bf{X}} \right) = \left\{ \begin{array}{l}
\max \left\{ {0,{g_v}\left( {\bf{X}} \right)} \right\},\quad v = 1,2, \ldots, p\\
\max \left\{ {0,\left| {{h_v}\left( {\bf{X}} \right) - \iota } \right|} \right\},\quad v = p + 1, \ldots, q
\end{array} \right..
\end{equation}

Leveraging (\ref{51515}) to (\ref{5621554145}), the state can be mathematically defined as:
\begin{equation}\label{4322341}
{\cal S} = \left\{ {s|s = \left( {cov,fea,div} \right)} \right\}.
\end{equation}
Building on this, the reward ${\cal R}$ can be expressed
\begin{equation}\label{1312323}
{\cal R} = \left( {cov,fea,div} \right) - \left( {cov',fea',div'} \right),
\end{equation}
where $\left( {cov',fea',div'} \right)$ represents the state of the next generated population. The record $Rec$, and experience replay $EP$ can be expressed as follows:
\begin{equation}\label{26564}
Rec = \left( {cov,fea,div,op,{\cal R},cov',fea',div'} \right),
\end{equation}
\begin{equation}\label{5452331123}
EP = \left\{ {Re{c_1},Re{c_2}, \ldots ,Re{c_{\left| {EP} \right|}}} \right\}.
\end{equation}

\subsubsection{Customization of DQN-based MOEA} The customization of DQN-based MOEA includes three aspects, i.e., sub-objective division, decision variable determination, and DRL application implementation aspects. More concretely, we first divide the optimization problem in (\ref{4874515}) into four subjectives, including minimizing $R{B_{error}}$, $ - {\varsigma _1}{\gamma _{secure}}$, ${\varsigma _2}AAoI$, and ${\varsigma _3}\mu _{UAV}^3$. We then specify the decision variables to be optimized as ${{P_{BS}}}$, ${{P_{raddar}}}$, ${\mu _{BS}}$, and ${\mu _{UAV}}$. At last, we train the DQN to perform operator selection rather than using fixed preset operators, because this allows for a dynamic and adaptive search process, tailored to the specific characteristics of the optimization problem and the evolving state of the population \cite{ming2024constrained}.

Based on the above analysis, we propose a DQN-based MOEA for optimizing sensing, secrecy communication, and computing performances of LAWNs, which is detailed in Algorithm 1. Algorithm 1 is executed at the BS and the relay UAV ends. As can be seen, at the first stage, we initialize the algorithm, including parameters, sub-objectives, decision variables set, and feasible solutions set. 
Then, in the second stage, we train the DQN to perform the operator selection. Specifically, we first randomly select an operator from (\ref{5145}) and use $F_{MOEA}$ to obtain offspring and next generation population. On this basis, we can determine the new state and reward based on (\ref{4322341}) and( \ref{1312323}), respectively, and train the DQN by updating the experience replay $EP$ via (\ref{5452331123}). 

In the third stage, the trained DQN is used as MOEA. Specifically, we feed the population state into the trained DQN to obtain Q-values, and then adaptively select the operator based on these values. Afterthat, the offspring and next generated population are determined by $F_{MOEA}$. Besides, the reward, new state, and record can be updated. Finally, when the looping ends, the DQN-based MOEA outputs the optimized solutions for the problem.

\begin{algorithm}[]
	\renewcommand{\algorithmicrequire}{\textbf{Input:}}
	\renewcommand{\algorithmicensure}{\textbf{Output:}}
	\caption{DQN-based MOEA for ISCC Performances Optimizing Algorithm}
	\label{algor1}
	\begin{algorithmic}[1]
		\REQUIRE $OP$: the formulated optimization problem, $F_{MOEA}$: MOEA algorithm, $f_{eval}$: the evaluation function of MOEA algorithm, $N_{eval}$: the number of function evaluation times, $\max_{eval}$: the maximum number of function evaluation times, $EP$: the experience replay of DQN, $EP_{num}$: the size of $EP$.
		\ENSURE $s_{OP}$: the solution of optimization problem.
        \\ \textbf{// Stage 1: Parameters initialization}
        \STATE Initialize sub-objects and decision variables with $OP$
        \STATE Initialize the population $POP$%, dimension ${kk}$ and size $len\left( {{\bf{qq}}} \right)$ of queries ${\bf{Qu}}$
        \STATE Initialize the evaluation function of MOEA $f_{eval}$
        \STATE Initialize the value of $N_{eval}$ as 1
        \STATE   \textbf{While} ${N_{eval}} < {\max _{eval}}$ \textbf{do}
        \\ \textbf{// Stage 2: DQN training}
        \STATE   \quad \textbf{For} $E{P_{num}} < \left| {EP} \right|$ \textbf{do}
        \STATE   \quad \quad Randomly select an operator from (\ref{5145})
        \STATE   \quad \quad Generate offspring by using $F_{MOEA}$
        \STATE   \quad \quad Select next generated population by using $F_{MOEA}$
        \STATE   \quad \quad Determine the reward and new state based on (\ref{4322341}) and (\ref{1312323})
        \STATE   \quad \quad Train the DQN by updating the experience replay $EP$ via (\ref{5452331123})
        \STATE   \quad \textbf{End}
        \\ \textbf{// Stage 3: Trained DQN implementation to MOEA}
        %\STATE   \quad\textbf{For} each decision variable \textbf{do}
        \STATE \quad Input the state of population to trained DQN and obtain the Q value
        \STATE \quad Adaptively select the operator based on Q value
        \STATE \quad Generate offspring set with adopted MOEA $F_{MOEA}$ and selected operator
        \STATE \quad Select the population for the next generation with adopted MOEA $F_{MOEA}$
        \STATE \quad Calculate the reward and new state based on (\ref{4322341}) and (\ref{1312323}), and make a record with (\ref{26564})
        \STATE \quad $N_{eval}= N_{eval} + 1$        \STATE \textbf{End}
        \STATE Return $s_{OP}$
\end{algorithmic}
\end{algorithm}

\subsection{Complexity and convergence property analysis}

For the proposed Algorithm 1, its complexity mainly includes the MOEA execution, DQN training, and population state calculation parts \cite{ming2024constrained}.
Specifically, let the size of population, the number of objectives, and optimized variables be ${N_{POP}}$, $m$, and ${N_{OV}}$, respectively. Then, the complexity of MOEA execution is between $o\left( {mN_{POP}^2} \right)$ and $o\left( {N_{POP}^3} \right)$. In addition, let the number of neurons of DQN be $N_{ner}$, the complexity of training DQN is $o\left( {4N_{ner}^2} \right)$. Furthermore, the complexity of population state calculation is $o\left( {m{N_{POP}}{N_{OV}}} \right)$. On this basis, the total complexity of Algorithm 1 is from $o\left( {mN_{POP}^2 + 4N_{ner}^2 + m{N_{POP}}{N_{OV}}} \right)$ to $o\left( {N_{POP}^3 + 4N_{ner}^2 + m{N_{POP}}{N_{OV}}} \right)$.

% As for the algorithm's convergence property, on the one hand, the DRL-based MOEA is gradient-free optimization algorithm, its convergence doesn't depend on the complexity of optimization objects and constraints. On the other hand, the looping ending condition is function evaluation reaching maximum times, since the maximum function evaluation value is finite, the proposed algorithm can be ensured to terminate and converge \footnote{The quality of convergence (i.e., whether it converges to a global optimum) may not be guaranteed, since it depends on the factors including key parameters of MOEA, e.g., population size, mutation rate, crossover rate etc.} at last.

\section{Simulation results and analysis}

\subsection{Parameter Configurations}
To evaluate the performance superiority of the proposed algorithm, this section conducts simulation experiments under various parameter settings. Specifically, the simulation settings are presented in Table \ref{parameters}. Besides, to demonstrate the effectiveness of the proposed algorithm, two baselines are adopted as follows.

1) Genetic Algorithm (GA): GA~\cite{sun2022resource} is a mature search algorithm inspired by biological evolution, which simulates selection, crossover, and mutation to find global optimal solutions. GA's robust global search capability makes it well-suited for optimization problems. The key steps include chromosome encoding, fitness evaluation, selection, and the application of crossover and mutation. The iterative process continues until a termination condition is met, driving the population towards better solutions. %\textcolor{blue}{This work employs an improved GA to enhance search efficiency.}

2) Improved Multi-operator Differential Evolution (IMODE): IMODE~\cite{sallam2020improved} is an enhanced differential evolution algorithm designed for unconstrained optimization. It leverages multiple Differential Evolution (DE) operators by dividing the initial population into diverse sub-populations, each evolving with a different mutation strategy. The subpopulations are dynamically adjusted based on solution quality and diversity. Besides, a linear population size reduction strategy and sequential quadratic programming local search are applied to further enhance its performance.

%3) PSO.

%1) MOEAD2WA \cite{jiao2021two}. This is a decomposition-based MOEA for highly constrained many-objective problems (CMOPs). It utilizes a novel constraint-handling technique employing two types of weight vectors , i.e., feasible and infeasible, dynamically reducing the latter during the search to guide solutions across infeasible regions while retaining the former to direct the population towards the Pareto front.

%2) MOEADCMT \cite{chu2024competitive}. This approach transforms the CMOPs into competing subtasks, allocating computational resources based on a reward system tied to the archive's update frequency. Furthermore, this approach incorporates a knowledge transfer mechanism, drawing parent solutions from both the primary and auxiliary tasks to improve the diversity.

%3) MOEADDAE \cite{zhu2020constrained}. This scheme integrates a detect-and-escape (DAE) strategy within the MOEA framework. The DAE strategy detects stagnation using the feasible ratio and constraint violation change rate, and adjusts the constraint violation weight to guide the search. This scheme also adopts an improved epsilon-constraint method with precisely defined start and stop criteria.

%3) MOEADD \cite{li2014evolutionary}.

\begin{table}[]
\centering
\caption{Simulation parameter settings.}
\label{parameters}
\begin{tabular}{|c|c|c|c|}
\hline
\textbf{Parameter} & \textbf{Value} & \textbf{Parameter} & \textbf{Value}                \\ \hline
$N_{nuser}$            & 4            & $di{s_{BS,UAV}}$       & 100m                      \\ \hline
$di{s_{UAV,user}}$            & 100m           & $dis_{ref}$       & 100m                  \\ \hline
$L_e$            & 2.4             & $P_{sum}$       & 2                  \\ \hline
$P_{UAV}$            & 1           & $P_{user}$       & 1                  \\ \hline
$T_{c,UAV}$             & 196            & $T_{c,BS}$         & 196                 \\ \hline
$T_{UAV}$            & 15              & $T_{user}$               & 4 \\ \hline
\end{tabular}
\end{table}

%\subsection{Convergence property evaluation}

% \subsection{Performance under different antenna number of BS}
\subsection{Performance vs. Number of Antennas}

\subsubsection{Overall Performance}
\begin{figure}[]
  \centering
  % Requires \usepackage{graphicx}
  \includegraphics[width=7.5cm]{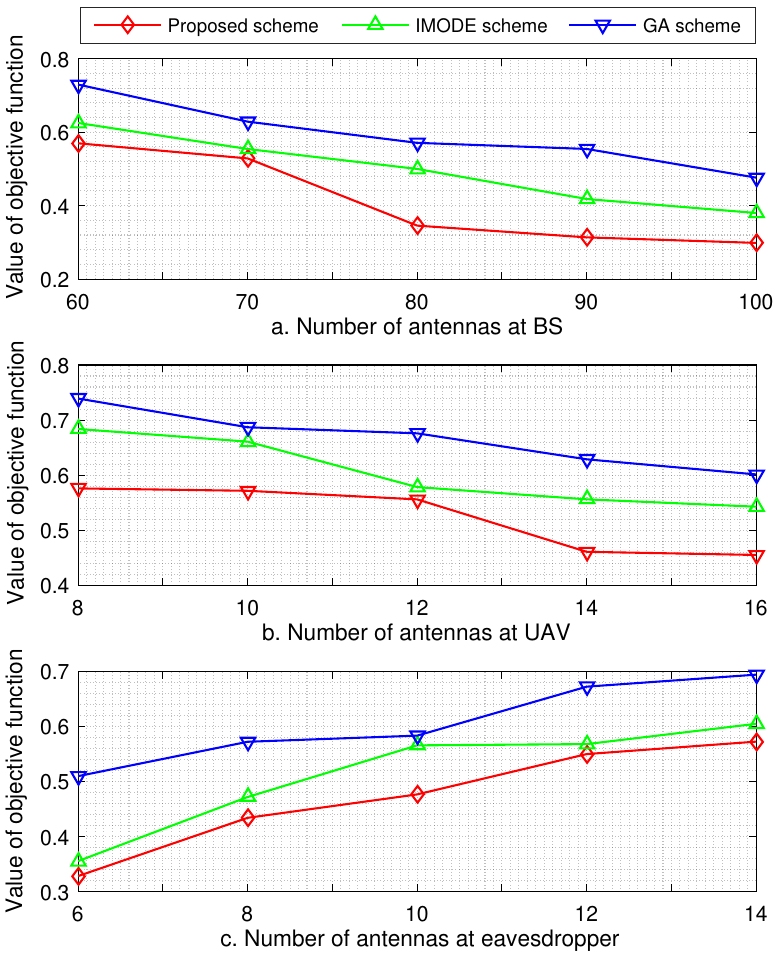}\\
  \caption{The objective function value under different antenna configurations.}
  \label{Ant}
\end{figure}

Figure~\ref{Ant} reports the objective value versus the number of antennas. For instance, in Fig.~\ref{Ant}a,  for the proposed scheme, we can see that as the number of BS antennas increases from 60 to 100, the function value decreases from 0.57 to 0.30. For comparison, IMODE drops from 0.63 to 0.38 and GA scheme from 0.73 to 0.48. This occurs because, without resource optimization in (\ref{4874515}), more antennas enhance sensing performance, yielding a smaller beampattern error. Besides that, more BS antennas also increases channel capacity, which, by (\ref{1415456123123}), raises the secrecy rate. This higher secrecy rate can drain the communication queue, leading to larger average AoI~\cite{kuang2020analysis}. To preserve AoI, the UAV’s computing rate needs to be increased, which in turn consumes more energy. Ultimately, jointly optimizing $P_{BS}$, $P_{sense}$, $\mu_{BS}$, and $\mu_{UAV}$ to balance the competing metrics yields a lower objective value as the number of antennas at BS increases.

Similarly, as illustrated in Fig.~\ref{Ant}b, increasing the number of antennas of the UAV from 8 to 16 reduces the objective function. For the proposed method, the value decreases from 0.58 to 0.46, corresponding to a reduction of approximately 21\%. For the comparison, IMODE decreases from 0.68 to 0.54 (about 21\% reduction) and GA from 0.74 to 0.60 (about 19\% reduction). The reason is that, given the fixed resource (i.e., BS's sum transmission power, and computation capacities of BS and UAV), increasing the number of UAV antennas can improve the anti-eavesdropping capacity, thereby enhancing secrecy rate. Similar to the previous analysis, higher secrecy rate indicates higher average AoI value and computational energy consumption of the UAV. Therefore, to minimize value of sub-objective functions jointly, the proposed scheme optimizes the resources $P_{BS}$, $P_{sens}$, $\mu_{BS}$ and $\mu_{UAV}$ and the final results indicate that the overall ISCC performance of LAWNs improves as the number of UAV antennas increases. The experimental results above demonstrate that increasing the number of antennas at either the base station or the UAV improves LAWNs' performance. Beyond this trend, we can also observe that the proposed method consistently attains lower objective values than the baselines across all configurations, showing its effectiveness for ISCC optimization in LAWNs.

In contrast to the previous findings, Fig.~\ref{Ant}c shows that increasing the number of antennas at the eavesdropper raises the objective value, indicating the performance decline. Specifically, when the eavesdropper antennas increase from 6 to 14, the objective value of the proposed scheme grows from 0.33 to 0.57. For comparison, that of IMODE increases from 0.36 to 0.60, while that of GA increases from 0.51 to 0.69. This trend arises because, under a fixed resource budget, additional eavesdropper antennas enlarge the eavesdropper’s channel capacity, thereby degrading secrecy-rate performance. Such a reduction in secrecy rate can further induce congestion in data transmission, which in turn increases the AoI. Together, these effects deteriorate the overall system performance. Nevertheless, the proposed method consistently achieves better performance than the baselines across all configurations, further validating its effectiveness and robustness.

% Accordingly, Fig. \ref{tu1} shows the sub-objective performances under different antenna number of BS, including beam-pattern error, secrecy rate, average AoI, and UAV computing's energy metrics from Fig. \ref{2} to Fig. \ref{5}. To de specific, in Fig. \ref{2}, the proposed scheme is not always better than baseline schemes, the reason is that the derived Pareto-optimal solution by the proposal  represents the solution where no single objective can be improved without degrading at least one other objective. Therefore, despite the proposed scheme cannot always performs best on single aspect, it can realize the best performance on the sum objective function, which is consistent with the optimization goal in Eq. \ref{4874515}.

% \begin{figure}[]
%   \centering
%   % Requires \usepackage{graphicx}
%   \includegraphics[width=6cm]{1.pdf}\\
%   \caption{The value of objective function under different antenna number of BS.}
%   \label{1}
% \end{figure}

\subsubsection{Sub-objective Performance}
\begin{figure*}[h]
\centering
\subfigure[]
{
   \label{2}
   %newdataave1
    \includegraphics[width=0.48\columnwidth]{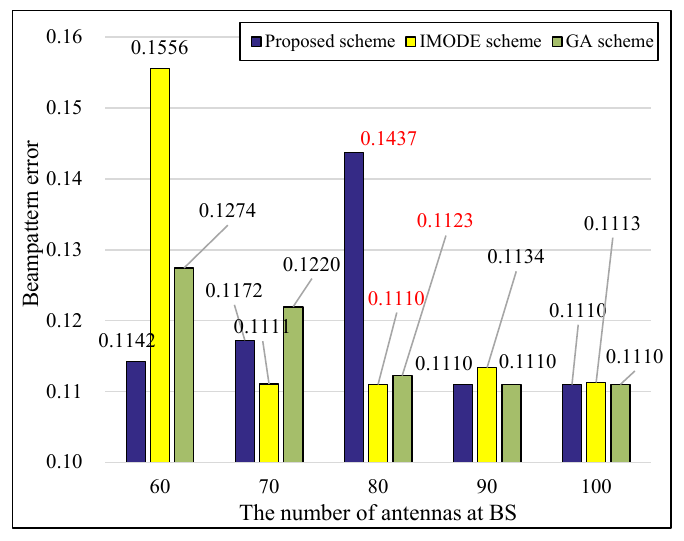}
}
\quad \quad
\hspace{-0.4in}
\subfigure[]
{
   \label{3}
   %newdataave1
    \includegraphics[width=0.48\columnwidth]{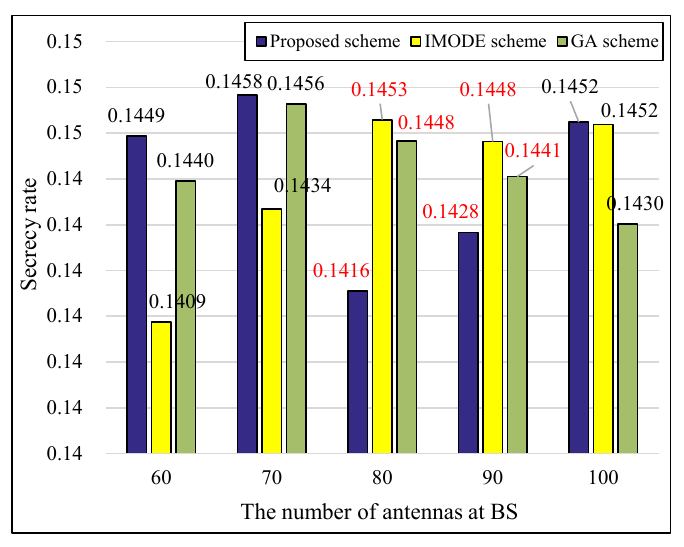}
}
\quad \quad
\hspace{-0.4in}
\subfigure[]
{
   \label{4}
   %newdataave1
    \includegraphics[width=0.48\columnwidth]{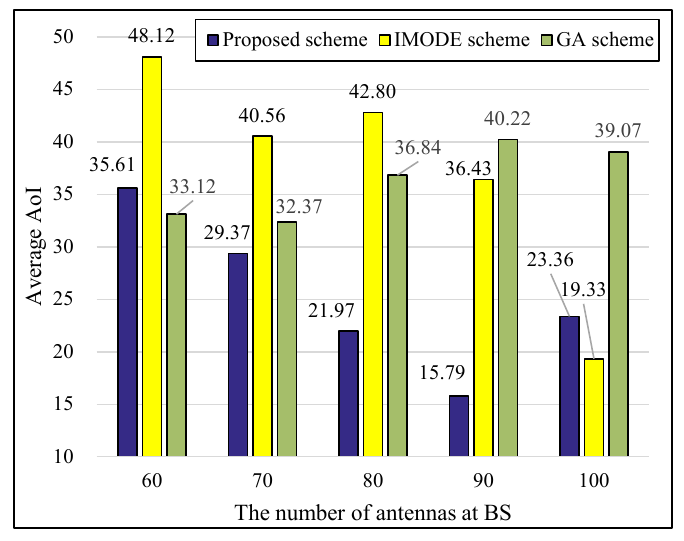}
}
 \quad \quad
\hspace{-0.4in}
\subfigure[]
{
   \label{5}
   %newdataave1
    \includegraphics[width=0.48\columnwidth]{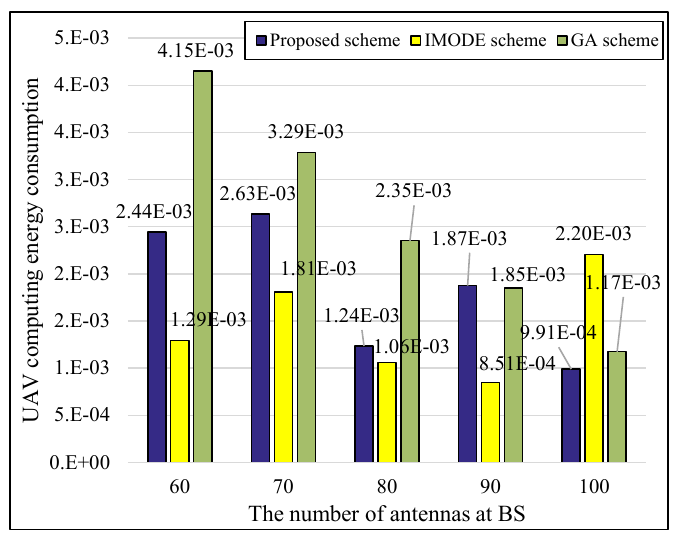}
}
\quad \quad
\hspace{-0.4in}

\caption{Performance of sub-objectives under various BS antenna configurations. (a) provides the beam-pattern error of sensing, (b) presents the secrecy rate of communications, (c) shows the average AoI, and (d) illustrated the computing energy consumption of the UAV.}

\label{BS_A}
\end{figure*}

\begin{figure*}[h]
\centering
\subfigure[]
{
   \label{7}
   %newdataave1
    \includegraphics[width=0.48\columnwidth]{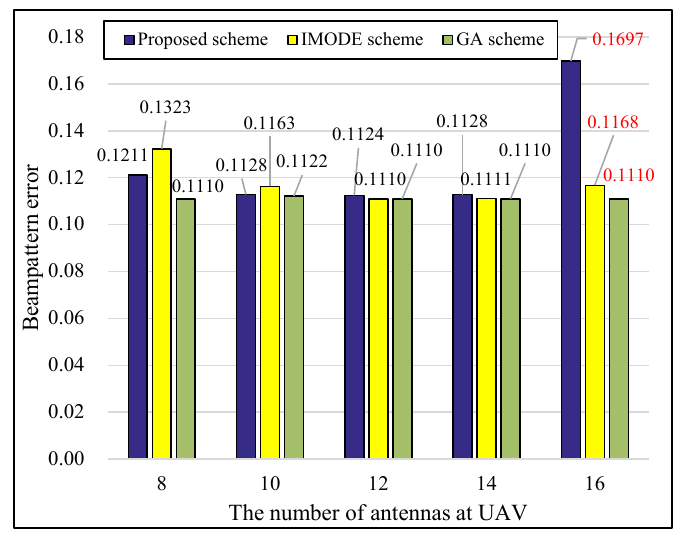}
}
\quad \quad
\hspace{-0.4in}
\subfigure[]
{
   \label{8}
   %newdataave1
    \includegraphics[width=0.48\columnwidth]{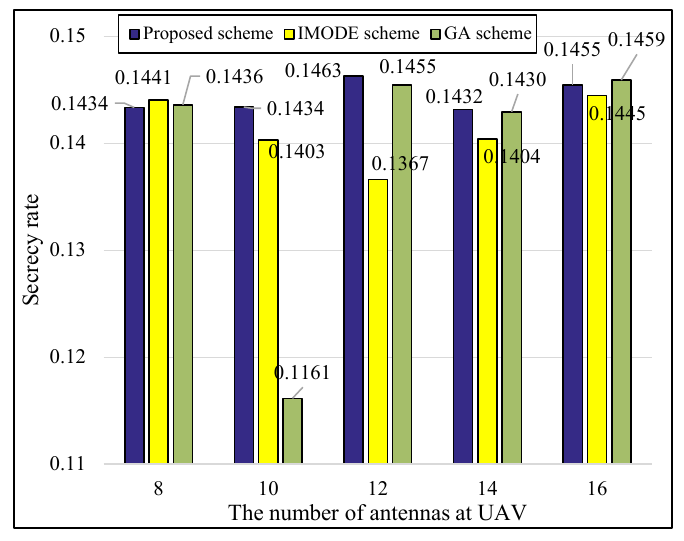}
}
\quad \quad
\hspace{-0.4in}
\subfigure[]
{
   \label{9}
   %newdataave1
    \includegraphics[width=0.48\columnwidth]{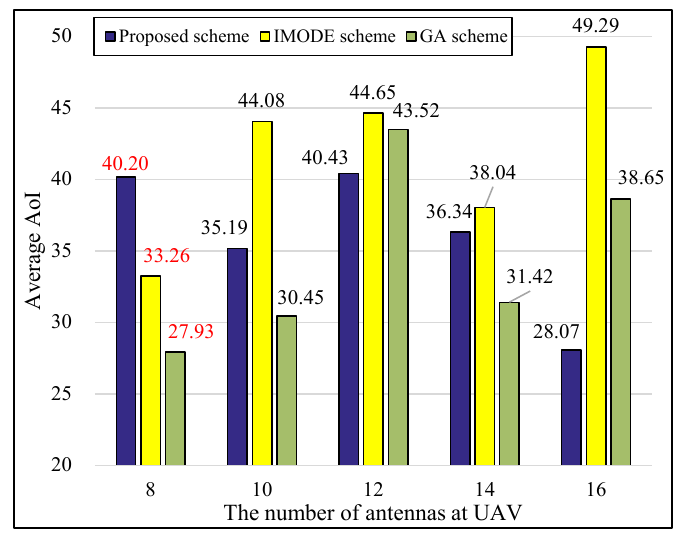}
}
 \quad \quad
\hspace{-0.4in}
\subfigure[]
{
   \label{xx}
   %newdataave1
    \includegraphics[width=0.48\columnwidth]{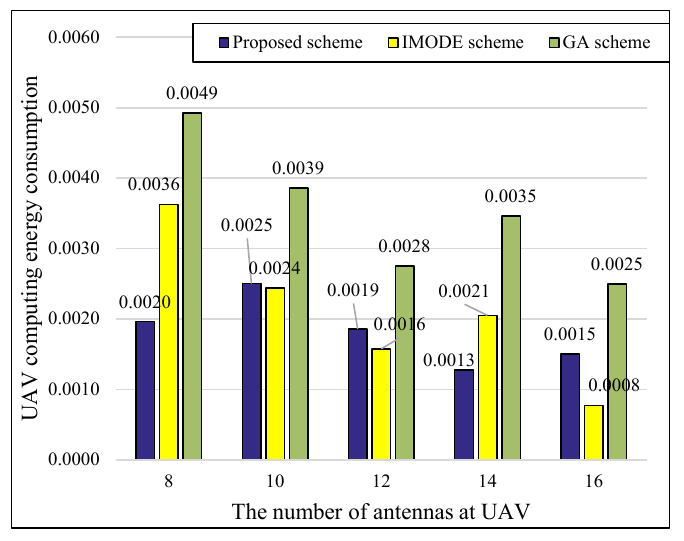}
}
\quad \quad
\hspace{-0.4in}

\caption{Performance of sub-objectives under various UAV antenna configurations. (a) provides the beam-pattern error of sensing, (b) presents the secrecy rate of communications, (c) shows the average AoI, and (d) illustrated the computing energy consumption of the UAV.}

\label{UAV_A}
\end{figure*}

\begin{figure*}[h]
\centering
\subfigure[]
{
   \label{12}
   %newdataave1
    \includegraphics[width=0.48\columnwidth]{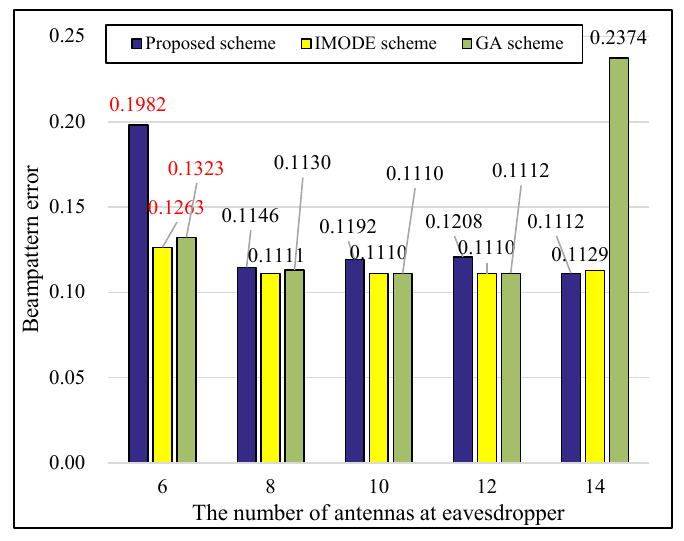}
}
\quad \quad
\hspace{-0.4in}
\subfigure[]
{
   \label{13}
   %newdataave1
    \includegraphics[width=0.48\columnwidth]{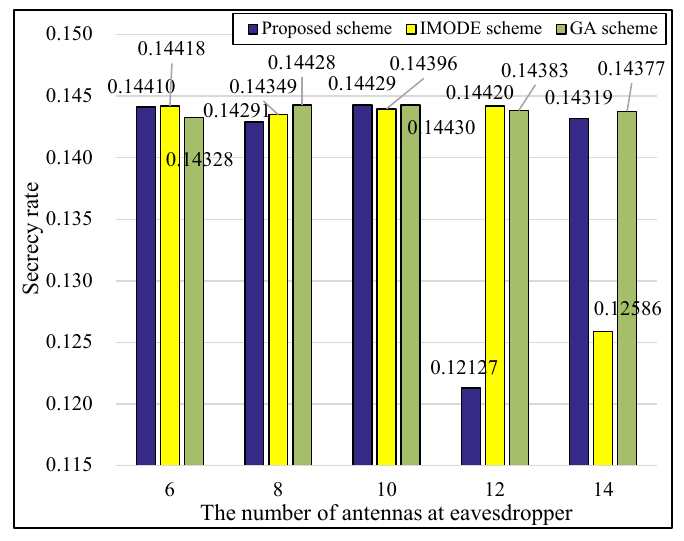}
}
\quad \quad
\hspace{-0.4in}
\subfigure[]
{
   \label{14}
   %newdataave1
    \includegraphics[width=0.48\columnwidth]{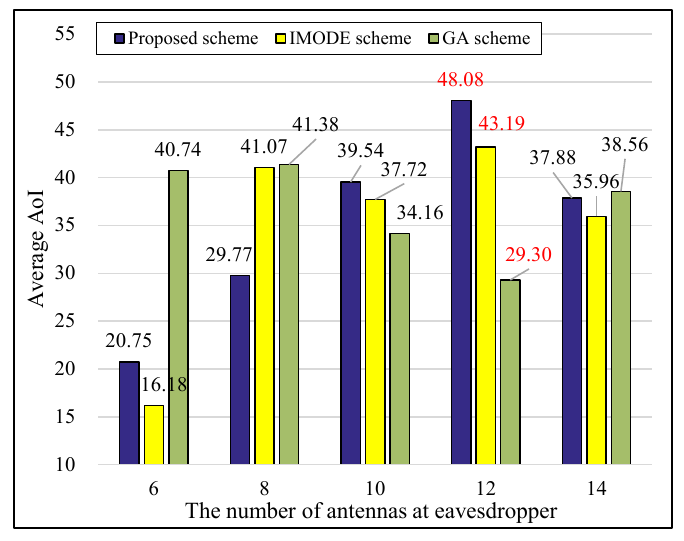}
}
 \quad \quad
\hspace{-0.4in}
\subfigure[]
{
   \label{xxx}
   %newdataave1
    \includegraphics[width=0.48\columnwidth]{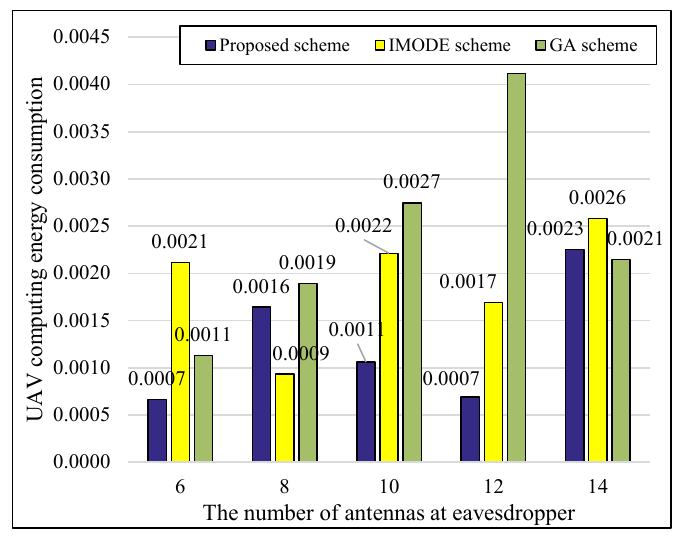}
}
\quad \quad
\hspace{-0.4in}

\caption{Performance of sub-objectives under various eavesdropper antenna configurations. (a) provides the beam-pattern error of sensing, (b) presents the secrecy rate of communications, (c) shows the average AoI, and (d) illustrated the computing energy consumption of the UAV.}

\label{EAV_A}
\end{figure*}

Building on the above, we further analyze the performance of the each sub-objective. Figs.~\ref{BS_A},~\ref{UAV_A}, and ~\ref{EAV_A} report values of sub-objectives, including beampattern error, secrecy rate, average AoI, and UAV computing energy consumption, under different antenna configurations. It can be observed that the proposed scheme does not uniformly dominate the baselines across all settings. Specifically, as shown in Fig.~\ref{BS_A}a, when the BS has 80 antennas, the beampattern error of the proposed scheme is 0.1437, higher than IMODE's 0.1110 and the GA's 0.1123. For secrecy rate, with 80 and 90 antennas of the BS \cite{meng2024network}, the proposed scheme attains 0.1416 and 0.1428, which are lower than IMODE's 0.1453 and 0.1448, and GA's 0.1448 and 0.1441.

Similar patterns can also be observed in Figs.~\ref{UAV_A} and~\ref{EAV_A}. Concretely, for UAV with 8 antennas, the AoI of the proposed scheme is 40.20, higher than IMODE's 33.26 and GA's 27.93. With 16 UAV antennas, the beampattern error of the proposed scheme is 0.1697, greater than IMODE's 0.1168 and GA's 0.1110. For the eavesdropper, with 6 antennas, the beampattern error of our scheme is 0.1982, which is higher than 0.1263 of IMODE and 0.1323 of GA. Furthermore, with 12 eavesdropper antennas, the AoI of our scheme is 48.08, higher than IMODE's 43.19 and GA's 29.30.

The analysis presented above indicate that, for the sub-objective, the proposed scheme cannot guarantee superiority over the baselines under every configuration. This is expected because the obtained solution is Pareto-optimal, which means no sub-objective can be improved without degrading at least one other sub-objective. Furthermore, the values of individual sub-objectives do not exhibit consistent trends with respect to the number of antennas. This is due to the multi-objective nature of the constructed problem, where the goal is to optimize the overall ISCC performance in LAWNs. In this context, fluctuations in a single sub-objective are compensated by the combined effect of other objectives, and as a result, no consistent pattern is observed at the individual level. In summary, while the proposed scheme may not dominate every metric individually, it consistently achieves optimal overall ISCC performance under various configurations, which aligns with the optimization goal defined in~(\ref{4874515}) and further confirms its effectiveness.

\subsection{Performance versus Noise Level}
\begin{figure}[]
  \centering
  % Requires \usepackage{graphicx}
  \includegraphics[width=7cm]{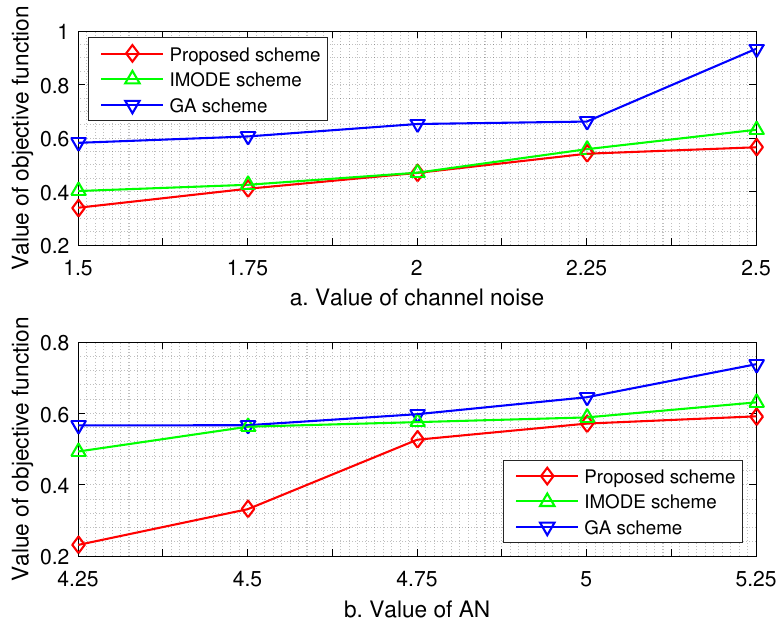}\\
  \caption{The objective function value under different noise configuration.}
  \label{Nos}
\end{figure}

Fig.~\ref{Nos} depicts the objective values under different noise conditions. As shown in Fig.~\ref{Nos}, for both channel noise and AN, increasing noise intensity significantly impacts overall system performance. Specifically, when the channel noise increases from 1.5 to 2.5, the objective value of the proposed schme rises from 0.34 to 0.57. Similarly, IMODE's value increases from 0.40 to 0.63, while GA's value increases from 0.58 to 0.93. This degradation occurs because higher channel noise directly impacts both sensing and communication, thereby weakening the overall system performance. A similar trend can be observed in Fig.~\ref{Nos}b. As the AN level increases from 4.25 to 5.25, the objective value of the proposed scheme rises from 0.23 to 0.59, while IMODE grows from 0.49 to 0.63 and GA increases from 0.57 to 0.74. This is attributed to the fact that inject AN degrade the performance (secrecy rate) of the eavesdropper, which leads to an increase in the objective function. Although increasing channel noise or AN deteriorate system performance, the proposed scheme consistently achieves the lowest objective value compared with the baselines, confirming its superior overall performance.

\begin{figure*}[h]
\centering
\subfigure[]
{
   \label{17}
   %newdataave1
    \includegraphics[width=0.48\columnwidth]{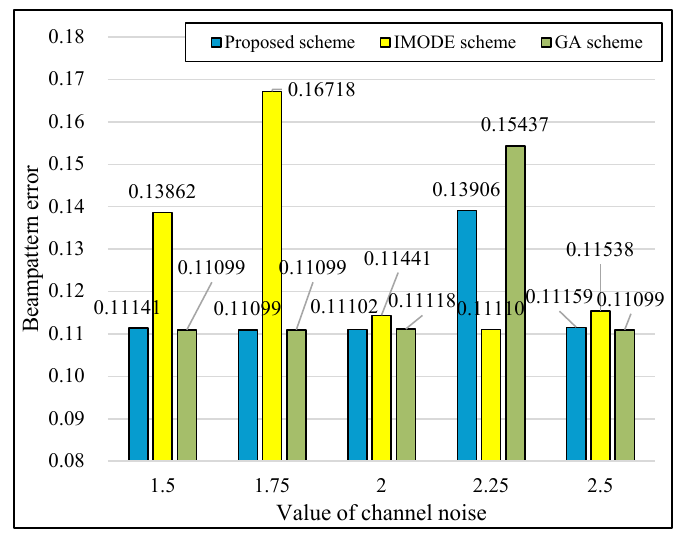}
}
\quad \quad
\hspace{-0.4in}
\subfigure[]
{
   \label{18}
   %newdataave1
    \includegraphics[width=0.48\columnwidth]{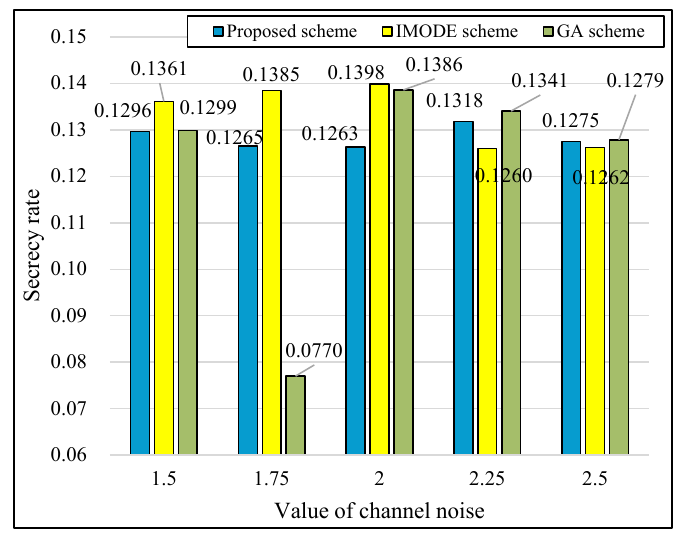}
}
\quad \quad
\hspace{-0.4in}
\subfigure[]
{
   \label{19}
   %newdataave1
    \includegraphics[width=0.48\columnwidth]{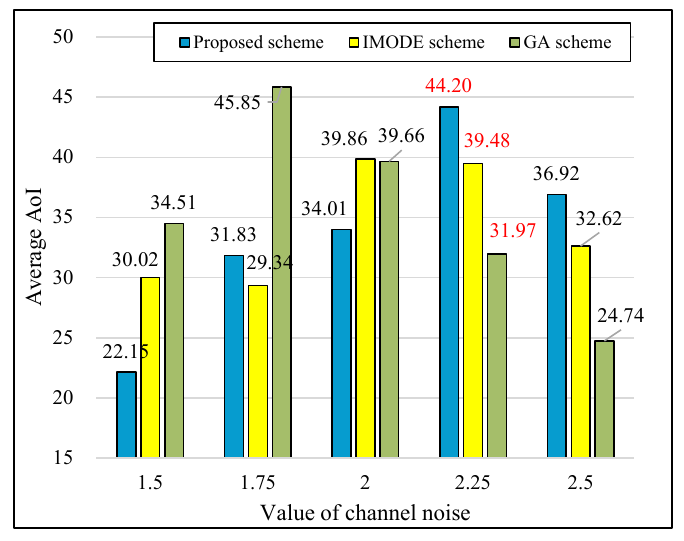}
}
 \quad \quad
\hspace{-0.4in}
\subfigure[]
{
   \label{20}
   %newdataave1
    \includegraphics[width=0.48\columnwidth]{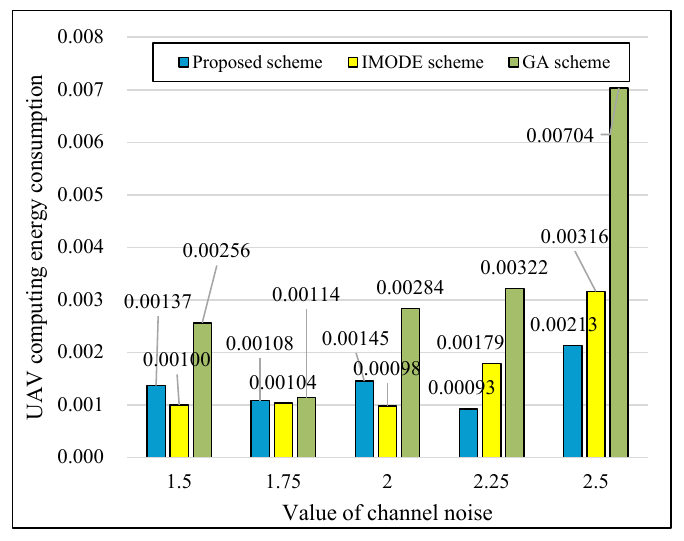}
}
\quad \quad
\hspace{-0.4in}

\caption{Performance of sub-objectives under various channel noise configurations. (a) provides the beam-pattern error of sensing, (b) presents the secrecy rate of communications, (c) shows the average AoI, and (d) illustrated the computing energy consumption of the UAV.}
\vspace{-0.15cm}
\label{CN}
\end{figure*}

\begin{figure*}[!ht]
\centering
\subfigure[]
{
   \label{22}
   %newdataave1
    \includegraphics[width=0.47\columnwidth]{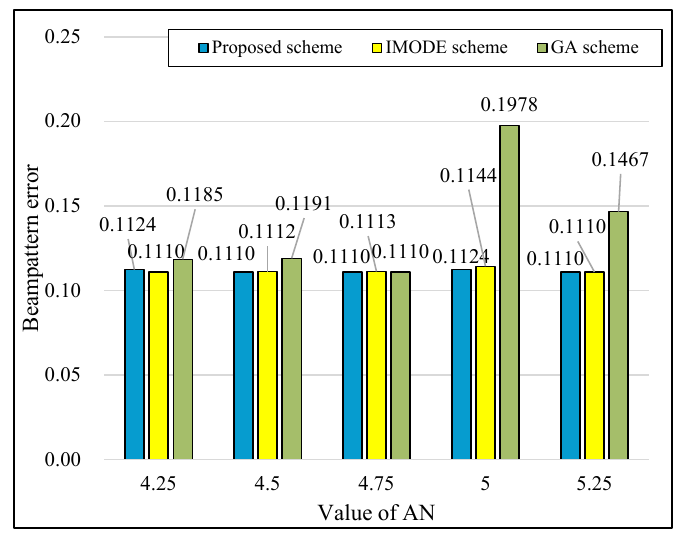}
}
\quad \quad
\hspace{-0.4in}
\subfigure[]
{
   \label{23}
   %newdataave1
    \includegraphics[width=0.47\columnwidth]{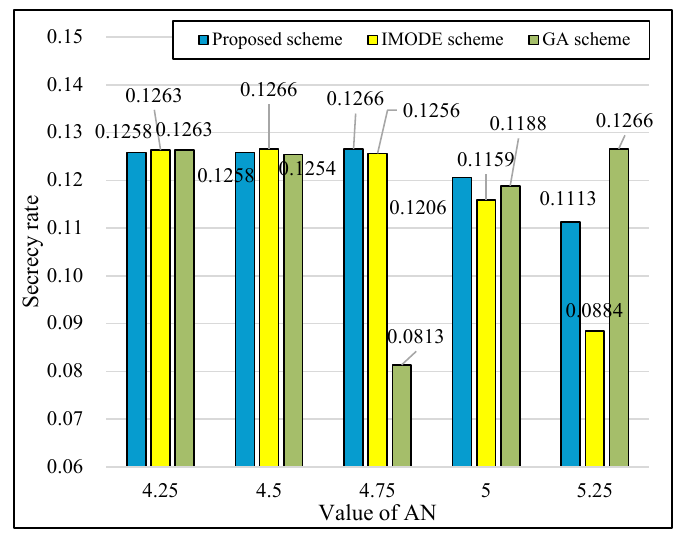}
}
\quad \quad
\hspace{-0.4in}
\subfigure[]
{
   \label{24}
   %newdataave1
    \includegraphics[width=0.47\columnwidth]{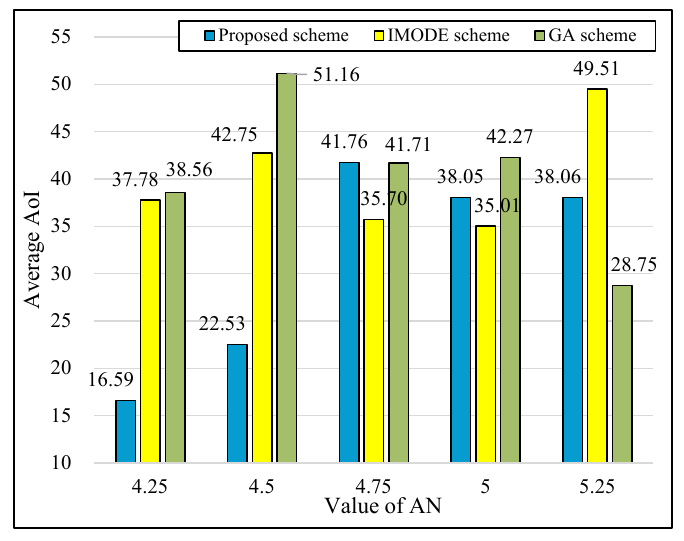}
}
 \quad \quad
\hspace{-0.4in}
\subfigure[]
{
   \label{25}
   %newdataave1
    \includegraphics[width=0.48\columnwidth]{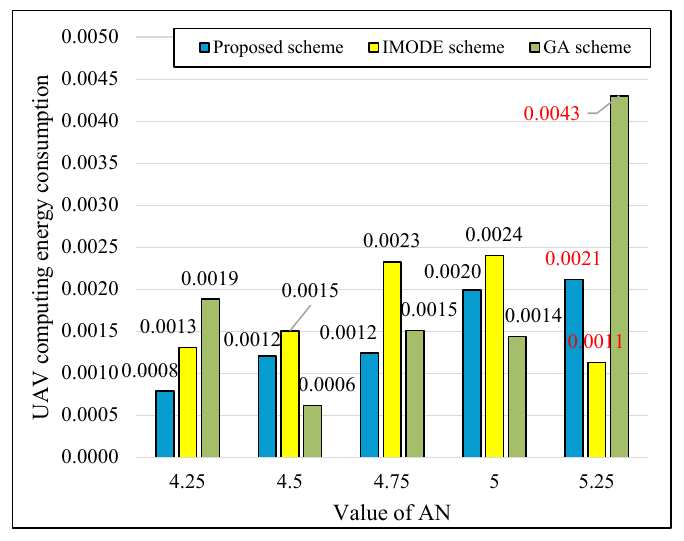}
}
\quad \quad
\hspace{-0.4in}

\caption{Performance of sub-objectives under various AN configurations. (a) provides the beam-pattern error of sensing, (b) presents the secrecy rate of communications, (c) shows the average AoI, and (d) illustrated the computing energy consumption of the UAV.}

\label{AN}
\end{figure*}

We further analyze the impact of noise on each sub-objective, and the results are illustrated in Figs.~\ref{CN} and \ref{AN}. Similar to the previous analysis, as the proposed scheme is Pareto-optimal, the proposed scheme cannot guarantee better performance than that of the baselines in every sub-objective. For example, when the channel noise level is 2.25, the AoI of the proposed scheme reaches 44.20, which is higher than IMODE's 39.48 and GA's 31.97. When AN level is 5.25, the UAV computing energy consumption is 0.0021, which is higher than IMODE's 0.0011, but lower than GA's 0.0043. Moreover, we observe that increasing channel noise does not increase the beampattern error, indicating that sensing performance remains relatively stable. This is because, when channel noise grows, the proposed scheme allocates more transmission power to sensing to maintain the beampattern error. As a trade-off, communication power and AN power are reduced, which in turn increases the AoI and slightly raises UAV energy consumption. In essence, the proposed scheme balances the impact of noise on sensing to the AoI and computational energy consumption through resource optimization, which helps ensure overall system performance and further demonstrates the advanced nature of the proposed scheme.

% \subsection{Performance vs. Different Levels of artificial noise}
% Fig. \ref{21} depicts the objective function's value under different values of artificial noise. Initially, increasing the artificial noise enhances the objective function. This occurs because the injected noise reduces the eavesdropper's secrecy rate, leading to an overall increase in the objective function. After optimizing resources for $P_{BS}$, $P_{sens}$, $\mu_{BS}$ and $\mu_{UAV}$, the relationship remains: increasing artificial noise continues to increase the objective function. Furthermore, our proposed scheme achieves the lowest objective function value compared to the listed baselines, indicating its superior performance.

% Performances of beam-pattern error, secrecy rate, average AoI, and UAV computing's energy are further illustrated in Fig. \ref{2121}. From Fig. \ref{22} to Fig. \ref{25} These figures highlight that the proposed scheme, which derives Pareto-optimal solutions, may not achieve the absolute best performance for every individual metric. Instead, it excels at optimizing the overall objective function  in Eq. \ref{4874515}, thereby achieving a well-balanced trade-off across all performance indicators.

% \begin{figure}[]
%   \centering
%   % Requires \usepackage{graphicx}
%   \includegraphics[width=6cm]{21.pdf}\\
%   \caption{The value of objective function under different values of artificial noise.}
%   \label{21}
% \end{figure}

\section{Conclusion}
In this paper, we have addressed a critical research gap by explicitly incorporating secrecy communication constraints into the ISCC performance optimization. This ensures that LAWNs remain secure against detection while still delivering high-quality sensing and computing services. We have derived beam pattern error, secrecy rate, and AoI to characterize ISCC performance and formulate a multi-objective optimization problem. Based on this, we have proposed a DQN-based multi-objective evolutionary algorithm to achieve the optimal solution. Through comprehensive simulations, we have demonstrated that the proposed algorithm achieves superior performance, outperforming state-of-the-art baseline schemes in jointly optimizing sensing accuracy, communication secrecy, and computation. This work highlights the significance of the security of the communication in LAWNs and offers a viable solution for reliable and efficient operations of LAWNs.
\vspace{-0.5cm}
\section*{Appendix A}
Based on \cite{kuang2020analysis}, the BS and UAV-based computing pipeline can be regarded as the $GI/M/1$ queue system for average AoI derivation. In particular, the BS's computing time $O_i$ and the data transmission time $Y_i$ are i.i.d. exponential random variables, which satisfy $\mathbb{E}\left[ {{O_i}} \right] = {1 \mathord{\left/
 {\vphantom {1 {{\mu _{BS}}}}} \right.
 \kern-\nulldelimiterspace} {{\mu _{BS}}}}$ and $\mathbb{E}\left[ {{Y_i}} \right] = {1 \mathord{\left/
 {\vphantom {1 {{\mu _{trans}}}}} \right.
 \kern-\nulldelimiterspace} {{\mu _{trans}}}}$, respectively. Therefore, we have
\begin{equation}\label{1514551}
\mathbb{E}\left[ {{B_i}} \right] = \mathbb{E}\left[ {{O_i} + {Y_i}} \right] = \mathbb{E}\left[ {{O_i}} \right] + \mathbb{E}\left[ {{Y_i}} \right] = \frac{1}{{{\mu _{BS}}}} + \frac{1}{{{\mu _{trans}}}},
\end{equation}
\begin{equation}\label{1514552}
\mathbb{E}\left[ {{B_i}{B_{i - 1}}} \right] = {\left( {\mathbb{E}\left[ B \right]} \right)^2} = \frac{1}{{\mu _{BS}^2}} + \frac{1}{{\mu _{trans}^2}} + \frac{2}{{{\mu _{BS}}{\mu _{trans}}}},
\end{equation}
and 
\begin{equation}\label{1514553}
\mathbb{E}\left[ {B_{i - 1}^2} \right] = \mathbb{E}\left[ {{B^2}} \right] = \frac{2}{{\mu _{BS}^2}} + \frac{2}{{\mu _{trans}^2}} + \frac{2}{{{\mu _{BS}}{\mu _{trans}}}}.
\end{equation}
% \begin{equation}\label{1514553}\small
% \mathbb{E}\left[ {B_{i - 1}^2} \right] = \mathbb{E}\left[ {{B^2}} \right] = \mathbb{E}\left[ {{{\left( {O + Y} \right)}^2}} \right] = \frac{2}{{\mu _{BS}^2}} + \frac{2}{{\mu _{trans}^2}} + \frac{2}{{{\mu _{BS}}{\mu _{trans}}}}.
% \end{equation}

For the $i$-th data packet, its waiting time $W{a_i}$ depends on the $(i-1)$-th data's waiting time ${T_{i - 1}}$ and the interval time $B_i$. Specifically, if the $i$-th data packet arrives at the relay UAV’s computing queue while the $(i-1)$-th data is still waiting or in service, then $W{a_i} = {T_{i - 1}} - {B_i}$. Otherwise, we have $W{a_i} = 0$. On this basis, the waiting time of $i$-th data packet is determined as
\begin{equation}\label{121253}
W{a_i} = {\left( {{T_{i - 1}} - {B_i}} \right)^ + }.
\end{equation}
Since ${T_i} = W{a_i} + S{e_i}$, we have:
\begin{align}\label{1543536}
\mathbb{E}\left[ {{T_i}{B_{i - 1}}} \right] &= \mathbb{E}\left[ {\left( {W{a_i} + S{e_i}} \right){B_{i - 1}}} \right]\\
 &= \mathbb{E}\left[ {W{a_i}{B_{i - 1}}} \right] + \mathbb{E}\left[ {S{e_i}{B_{i - 1}}} \right].
\end{align}
In this regard, the waiting time $W{a_i}$ can be calculated as
\begin{align}\label{1556523232123}
W{a_i} &= {\left( {{T_{i - 1}} - {B_i}} \right)^ + } \\ \notag
&= {\left( {W{a_{i - 1}} + S{e_{i - 1}} - {B_i}} \right)^ + }\\ \notag
&= {\left( {{{\left( {{T_{i - 2}} - {B_{i - 1}}} \right)}^ + } + S{e_{i - 1}} - {B_i}} \right)^ + },
\end{align}
where the system time ${T_{i - 2}}$ is independent of ${S{e_{i - 1}}}$, ${{B_{i - 1}}}$ and $B_i$. When the queue is steady, the system time is stochastically identical, indicating $T\mathop  = \limits^{st} {T_i}\mathop  = \limits^{st} {T_{i - 1}}\mathop  = \limits^{st} {T_{i - 2}}$. Based on \cite{kuang2020analysis}, the probability density function (PDF) of $T_i$ in the $GI/M/1$ queue system is
\begin{equation}\label{1532312313}
{f_T}\left( t \right) = {\mu _{UAV}}\left( {1 - \upsilon } \right){e^{ - {\mu _{UAV}}\left( {1 - \upsilon } \right)t}},t \ge 0,
\end{equation}
where $\upsilon$ can be denoted as
\begin{equation}\label{1513213}
\upsilon  = {\kappa ^*}\left( {{\mu _{UAV}}\left( {1 - \upsilon } \right)} \right),
\end{equation}
and ${\kappa ^*}\left( \cdot \right)$ is the Laplace-Stieltjes transform operation. In addition, according to \cite{florescu2014probability}, the PDF of $B_i$ is
\begin{equation}\label{1662447}
{f_B}\left( \kappa  \right) = \frac{{{\mu _{trans}}{\mu _{BS}}}}{{{\mu _{trans}} - {\mu _{BS}}}}\left( {{e^{ - {\mu _{BS}}\kappa }} - {e^{ - {\mu _{trans}}\kappa }}} \right),\kappa  > 0.
\end{equation}

On this basis, we have
\begin{align}\label{63545}
{\kappa ^*}\left( s \right) &= \mathbb{E}\left[ {{e^{ - sB}}} \right] = \int_0^\infty  {{f_B}\left( \kappa  \right)} {e^{ - s\kappa }}d\kappa \\ \notag
 &= \frac{{{\mu _{trans}}{\mu _{BS}}}}{{{\mu _{trans}} - {\mu _{BS}}}}\left( {\frac{1}{{{\mu _{BS}} + s}} - \frac{1}{{{\mu _{trans}} + s}}} \right).
\end{align}
By combing (\ref{1513213}) and (\ref{63545}), we have
\begin{equation}\label{48844485525}
\begin{array}{l}
\upsilon  = {\kappa ^*}\left( {{\mu _{UAV}}\left( {1 - \upsilon } \right)} \right)\\
 = \frac{{{\mu _{trans}}{\mu _{BS}}}}{{{\mu _{trans}} - {\mu _{BS}}}}\left( {\frac{1}{{{\mu _{BS}} + {\mu _{UAV}}\left( {1 - \upsilon } \right)}} - \frac{1}{{{\mu _{trans}} + {\mu _{UAV}}\left( {1 - \upsilon } \right)}}} \right)
\end{array}.
\end{equation}

For ease of presentation, let $\omega  = {\mu _{UAV}}\left( {1 - \upsilon } \right)$, and then the PDF of $T$ can be written as
\begin{equation}\label{9525}
{f_T}\left( t \right) = \omega {e^{ - \omega t}},t \ge 0.
\end{equation}
Under the condition ${{B_{i - 1}} = {\kappa _i}}$, we have (\ref{595615145}) and (\ref{26155414545}).

\newcounter{mytempeqncnt}
%\hrulefill %这里有一条线，如果你想要
\begin{figure*}[]%公式位置按图放置调整
\small

\begin{align}\label{595615145}
\mathbb{E}\left[ {W{a_i}|{B_{i - 1}} = {\kappa _i}} \right] &= \mathbb{E}\left[ {{{\left( {{{\left( {{T_{i - 2}} - {\kappa _i}} \right)}^ + } + S{e_{i - 1}} - {B_i}} \right)}^ + }|{B_{i - 1}} = {\kappa _i}} \right]  = \mathbb{E}\left[ {{{\left( {{{\left( {{T_{i - 2}} - {\kappa _i}} \right)}^ + } + S{e_{i - 1}} - {B_i}} \right)}^ + }} \right]\\ \notag
 &= \int_0^\infty  {{f_T}\left( t \right)} \int_0^\infty  {{f_S}\left( s \right)} \int_0^\infty  {{f_B}\left( \kappa  \right)} \left( {{{\left( {{{\left( {t - {\kappa _i}} \right)}^ + } + s - \kappa } \right)}^ + }} \right)dtdsd\kappa \\ \notag
 &= \frac{{{\varpi _1} - 1}}{{{\mu _{BS}}}} - \frac{{{\varpi _2} + 1}}{{{\mu _{trans}}}} + \frac{1}{{{\mu _{UAV}}}}  
 + \left( {\frac{1}{{{\varpi _3}}} - \frac{{{\varpi _1}}}{{{\mu _{BS}} + {\varpi _3}}} + \frac{{{\varpi _2}}}{{{\mu _{trans}} + {\varpi _3}}}} \right){e^{ - {\varpi _3}{\kappa _i}}}
\end{align}

\begin{align}\label{26155414545}
\mathbb{E}\left[ {W{a_i}{B_{i - 1}}} \right] &= \int_0^\infty  {{\kappa _i}}\mathbb{E} \left[ {W{a_i}|{B_{i - 1}} = {\kappa _i}} \right]{f_{{B_i}}}\left( {{\kappa _i}} \right)d{\kappa _i} = \frac{{{\mu _{trans}} + {\mu _{BS}}}}{{{\mu _{trans}}{\mu _{BS}}}}\left( {\frac{1}{{{\mu _{UAV}}}} - \frac{{{\varpi _1} - 1}}{{{\mu _{BS}}}} - \frac{{{\varpi _2} + 1}}{{{\mu _{trans}}}}} \right)\\ \notag
 &+ \frac{{{\mu _{trans}}{\mu _{BS}}}}{{{\mu _{trans}} - {\mu _{BS}}}}\left( {\frac{1}{{{{\left( {{\mu _{BS}} + {\varpi _3}} \right)}^2}}} - \frac{1}{{{{\left( {{\mu _{trans}} + {\varpi _3}} \right)}^2}}}} \right)\left( {\frac{1}{{{\varpi _3}}} - \frac{{{\varpi _1}}}{{{\mu _{BS}} + {\varpi _3}}} + \frac{{{\varpi _2}}}{{{\mu _{trans}} + {\varpi _3}}}} \right)
\end{align}

\hrulefill % %这里有一条线，如果你想要
%\vspace*{4pt} %留空白，可自己调整
\end{figure*}

%\begin{equation}\label{595615145}\small
%\begin{array}{l}
%\mathbb{E}\left[ {W{a_i}|{B_{i - 1}} = {\kappa _i}} \right]\\
% = \mathbb{E}\left[ {{{\left( {{{\left( {{T_{i - 2}} - {\kappa _i}} \right)}^ + } + S{e_{i - 1}} - {B_i}} \right)}^ + }|{B_{i - 1}} = {\kappa _i}} \right]\\
% = \mathbb{E}\left[ {{{\left( {{{\left( {{T_{i - 2}} - {\kappa _i}} \right)}^ + } + S{e_{i - 1}} - {B_i}} \right)}^ + }} \right]\\
% = \int_0^\infty  {{f_T}\left( t \right)} \int_0^\infty  {{f_S}\left( s \right)} \int_0^\infty  {{f_B}\left( \kappa  \right)} \left( {{{\left( {{{\left( {t - {\kappa _i}} \right)}^ + } + s - \kappa } \right)}^ + }} \right)dtdsd\kappa \\
% = \frac{{{\varpi _1} - 1}}{{{\mu _{BS}}}} - \frac{{{\varpi _2} + 1}}{{{\mu _{trans}}}} + \frac{1}{{{\mu _{UAV}}}} \\ 
% + \left( {\frac{1}{{{\varpi _3}}} - \frac{{{\varpi _1}}}{{{\mu _{BS}} + {\varpi _3}}} + \frac{{{\varpi _2}}}{{{\mu _{trans}} + {\varpi _3}}}} \right){e^{ - {\varpi _3}{\kappa _i}}}
%\end{array}.
%\end{equation}

Since the interval time ${{B_{i - 1}}}$ is independent of $Se_i$, we have
\begin{equation}\label{452625}
\mathbb{E}\left[ {{T_i}{B_{i - 1}}} \right] = \mathbb{E}\left[ {W{a_i}{B_{i - 1}}} \right] + \mathbb{E}\left[ {S{e_i}} \right]\mathbb{E}\left[ {{B_{i - 1}}} \right],
\end{equation}
where $\mathbb{E}\left[ {S{e_i}} \right] = {1 \mathord{\left/
 {\vphantom {1 {{\mu _{UAV}}}}} \right.
 \kern-\nulldelimiterspace} {{\mu _{UAV}}}}$. Finally, based on (\ref{21212}), (\ref{1514552}), (\ref{1514553}), (\ref{595615145}), and (\ref{26155414545}), we obtain (\ref{456123145456452}), and therefore Theorem 1 is proved.
% \vspace{-0.2cm}
\bibliography{ref}{}

% Generated by IEEEtran.bst, version: 1.14 (2015/08/26)
\begin{thebibliography}{10}
\providecommand{\url}[1]{#1}
\csname url@samestyle\endcsname
\providecommand{\newblock}{\relax}
\providecommand{\bibinfo}[2]{#2}
\providecommand{\BIBentrySTDinterwordspacing}{\spaceskip=0pt\relax}
\providecommand{\BIBentryALTinterwordstretchfactor}{4}
\providecommand{\BIBentryALTinterwordspacing}{\spaceskip=\fontdimen2\font plus
\BIBentryALTinterwordstretchfactor\fontdimen3\font minus \fontdimen4\font\relax}
\providecommand{\BIBforeignlanguage}[2]{{%
\expandafter\ifx\csname l@#1\endcsname\relax
\typeout{** WARNING: IEEEtran.bst: No hyphenation pattern has been}%
\typeout{** loaded for the language `#1'. Using the pattern for}%
\typeout{** the default language instead.}%
\else
\language=\csname l@#1\endcsname
\fi
#2}}
\providecommand{\BIBdecl}{\relax}
\BIBdecl

\bibitem{xie2025joint}
W.~Xie, G.~Sun, B.~Liu, J.~Li, J.~Wang, H.~Du, D.~Niyato, and D.~I. Kim, ``Joint optimization of uav-carried irs for urban low altitude mmwave communications with deep reinforcement learning,'' \emph{arXiv preprint arXiv:2501.02787}, 2025.

\bibitem{jepsen2024survey}
J.~H. Jepsen, K.~H. Laursen, and K.~Jensen, ``A survey of state-of-the-art u-space research,'' in \emph{2024 10th International Conference on Automation, Robotics and Applications (ICARA)}.\hskip 1em plus 0.5em minus 0.4em\relax IEEE, 2024, pp. 265--272.

\bibitem{wang2025toward}
Y.~Wang, G.~Sun, Z.~Sun, J.~Wang, J.~Li, C.~Zhao, J.~Wu, S.~Liang, M.~Yin, P.~Wang \emph{et~al.}, ``Toward realization of low-altitude economy networks: Core architecture, integrated technologies, and future directions,'' \emph{arXiv preprint arXiv:2504.21583}, 2025.

\bibitem{zhao2025generative}
C.~Zhao, J.~Wang, R.~Zhang, D.~Niyato, G.~Sun, H.~Du, D.~I. Kim, and A.~Jamalipour, ``Generative ai-enabled wireless communications for robust low-altitude economy networking,'' \emph{arXiv preprint arXiv:2502.18118}, 2025.

\bibitem{wang2025safeguarding}
J.~Wang, J.~He, G.~Sun, Z.~Xiong, D.~Niyato, S.~Mao, D.~I. Kim, and T.~Xiang, ``Safeguarding isac performance in low-altitude wireless networks under channel access attack,'' \emph{arXiv preprint arXiv:2508.15838}, 2025.

\bibitem{yuan2025ground}
W.~Yuan, Y.~Cui, J.~Wang, F.~Liu, G.~Sun, T.~Xiang, J.~Xu, S.~Jin, D.~Niyato, S.~Coleri \emph{et~al.}, ``From ground to sky: Architectures, applications, and challenges shaping low-altitude wireless networks,'' \emph{arXiv preprint arXiv:2506.12308}, 2025.

\bibitem{he2025satellite}
S.~He, J.~Wang, Y.-C. Liang, G.~Sun, and D.~Niyato, ``Satellite-assisted low-altitude economy networking: Concepts, applications, and opportunities,'' \emph{arXiv preprint arXiv:2505.04098}, 2025.

\bibitem{zhao2025temporal}
C.~Zhao, R.~Zhang, J.~Wang, D.~Niyato, G.~Sun, H.~Du, Z.~Li, A.~Jamalipour, and D.~I. Kim, ``Temporal spectrum cartography in low-altitude economy networks: A generative ai framework with multi-agent learning,'' \emph{arXiv preprint arXiv:2505.15571}, 2025.

\bibitem{cai2025secure}
L.~Cai, J.~Wang, R.~Zhang, Y.~Zhang, T.~Jiang, D.~Niyato, X.~Wang, A.~Jamalipour, and X.~Shen, ``Secure physical layer communications for low-altitude economy networking: A survey,'' \emph{arXiv preprint arXiv:2504.09153}, 2025.

\bibitem{wen2024survey}
D.~Wen, Y.~Zhou, X.~Li, Y.~Shi, K.~Huang, and K.~B. Letaief, ``A survey on integrated sensing, communication, and computation,'' \emph{IEEE Communications Surveys \& Tutorials}, 2024.

\bibitem{wang2024generative}
J.~Wang, H.~Du, D.~Niyato, J.~Kang, S.~Cui, X.~Shen, and P.~Zhang, ``Generative ai for integrated sensing and communication: Insights from the physical layer perspective,'' \emph{IEEE Wireless Communications}, vol.~31, no.~5, pp. 246--255, 2024.

\bibitem{zhang2024cooperative}
Y.~Zhang, H.~Shan, Y.~Zhou, Z.~Shi, L.~Sheng, and Y.~Liu, ``Cooperative beamforming design for anti-uav isac systems,'' \emph{IEEE Transactions on Wireless Communications}, 2024.

\bibitem{jing2024isac}
X.~Jing, F.~Liu, C.~Masouros, and Y.~Zeng, ``Isac from the sky: Uav trajectory design for joint communication and target localization,'' \emph{IEEE Transactions on Wireless Communications}, vol.~23, no.~10, pp. 12\,857--12\,872, 2024.

\bibitem{sun2024joint}
G.~Sun, L.~He, Z.~Sun, Q.~Wu, S.~Liang, J.~Li, D.~Niyato, and V.~C. Leung, ``Joint task offloading and resource allocation in aerial-terrestrial uav networks with edge and fog computing for post-disaster rescue,'' \emph{IEEE Transactions on Mobile Computing}, vol.~23, no.~9, pp. 8582--8600, 2024.

\bibitem{costanzo2020dynamic}
F.~Costanzo, P.~Di~Lorenzo, and S.~Barbarossa, ``Dynamic resource optimization and altitude selection in uav-based multi-access edge computing,'' in \emph{ICASSP 2020-2020 IEEE International Conference on Acoustics, Speech and Signal Processing (ICASSP)}.\hskip 1em plus 0.5em minus 0.4em\relax IEEE, 2020, pp. 4985--4989.

\bibitem{mei20223d}
H.~Mei, K.~Yang, Q.~Liu, and K.~Wang, ``3d-trajectory and phase-shift design for ris-assisted uav systems using deep reinforcement learning,'' \emph{IEEE Transactions on Vehicular Technology}, vol.~71, no.~3, pp. 3020--3029, 2022.

\bibitem{li2024unauthorized}
Z.~Li, Z.~Gao, K.~Wang, Y.~Mei, C.~Zhu, L.~Chen, X.~Wu, and D.~Niyato, ``Unauthorized uav countermeasure for low-altitude economy: Joint communications and jamming based on mimo cellular systems,'' \emph{IEEE Internet of Things Journal}, 2024.

\bibitem{atoev2019secure}
S.~Atoev, O.-J. Kwon, C.-Y. Kim, S.-H. Lee, Y.-R. Choi, and K.-R. Kwon, ``The secure uav communication link based on otp encryption technique,'' in \emph{2019 eleventh international conference on ubiquitous and future networks (ICUFN)}.\hskip 1em plus 0.5em minus 0.4em\relax IEEE, 2019, pp. 1--3.

\bibitem{sun2019physical}
X.~Sun, D.~W.~K. Ng, Z.~Ding, Y.~Xu, and Z.~Zhong, ``Physical layer security in uav systems: Challenges and opportunities,'' \emph{IEEE Wireless Communications}, vol.~26, no.~5, pp. 40--47, 2019.

\bibitem{tan2023secure}
H.~Tan, W.~Zheng, and P.~Vijayakumar, ``Secure and efficient authenticated key management scheme for uav-assisted infrastructure-less iovs,'' \emph{IEEE Transactions on Intelligent Transportation Systems}, vol.~24, no.~6, pp. 6389--6400, 2023.

\bibitem{xie2024joint}
H.~Xie, T.~Zhang, X.~Xu, D.~Yang, and Y.~Liu, ``Joint sensing, communication, and computation in uav-assisted systems,'' \emph{IEEE Internet of Things Journal}, vol.~11, no.~18, pp. 29\,412--29\,426, 2024.

\bibitem{li2023adaptive}
B.~Li, W.~Liu, W.~Xie, N.~Zhang, and Y.~Zhang, ``Adaptive digital twin for uav-assisted integrated sensing, communication, and computation networks,'' \emph{IEEE Transactions on Green Communications and Networking}, vol.~7, no.~4, pp. 1996--2009, 2023.

\bibitem{deng2024integrated}
C.~Deng, X.~Fang, and X.~Wang, ``Integrated sensing, communication, and computation with adaptive dnn splitting in multi-uav networks,'' \emph{IEEE Transactions on Wireless Communications}, 2024.

\bibitem{zhou2023uav}
Y.~Zhou, X.~Liu, X.~Zhai, Q.~Zhu, and T.~S. Durrani, ``Uav-enabled integrated sensing, computing, and communication for internet of things: Joint resource allocation and trajectory design,'' \emph{IEEE Internet of Things Journal}, vol.~11, no.~7, pp. 12\,717--12\,727, 2023.

\bibitem{chen2024uav}
J.~Chen, Y.~Xu, D.~Yang, and T.~Zhang, ``Uav-assisted iscc networks: Joint resource and trajectory optimization,'' \emph{IEEE Wireless Communications Letters}, vol.~13, no.~9, pp. 2372--2376, 2024.

\bibitem{zhou2019uav}
X.~Zhou, Q.~Wu, S.~Yan, F.~Shu, and J.~Li, ``Uav-enabled secure communications: Joint trajectory and transmit power optimization,'' \emph{IEEE Transactions on Vehicular Technology}, vol.~68, no.~4, pp. 4069--4073, 2019.

\bibitem{zhang2020uav}
Y.~Zhang, Z.~Mou, F.~Gao, J.~Jiang, R.~Ding, and Z.~Han, ``Uav-enabled secure communications by multi-agent deep reinforcement learning,'' \emph{IEEE Transactions on Vehicular Technology}, vol.~69, no.~10, pp. 11\,599--11\,611, 2020.

\bibitem{wang2021robust}
W.~Wang, X.~Li, R.~Wang, K.~Cumanan, W.~Feng, Z.~Ding, and O.~A. Dobre, ``Robust 3d-trajectory and time switching optimization for dual-uav-enabled secure communications,'' \emph{IEEE Journal on Selected Areas in Communications}, vol.~39, no.~11, pp. 3334--3347, 2021.

\bibitem{ding2022online}
Y.~Ding, Y.~Feng, W.~Lu, S.~Zheng, N.~Zhao, L.~Meng, A.~Nallanathan, and X.~Yang, ``Online edge learning offloading and resource management for uav-assisted mec secure communications,'' \emph{IEEE Journal of Selected Topics in Signal Processing}, vol.~17, no.~1, pp. 54--65, 2022.

\bibitem{tian2023uav}
W.~Tian, X.~Ding, G.~Liu, Y.~Dai, and Z.~Han, ``A uav-assisted secure communication system by jointly optimizing transmit power and trajectory in the internet of things,'' \emph{IEEE Transactions on Green Communications and Networking}, vol.~7, no.~4, pp. 2025--2037, 2023.

\bibitem{wang2024joint}
D.~Wang, D.~Wen, Y.~He, Q.~Chen, G.~Zhu, and G.~Yu, ``Joint device scheduling and resource allocation for iscc-based multi-view-multi-task inference,'' \emph{IEEE Internet of Things Journal}, 2024.

\bibitem{da2023privacy}
I.~W. da~Silva, D.~P. Osorio, and M.~Juntti, ``Privacy performance of mimo dual-functional radar-communications with internal adversary,'' in \emph{2023 IEEE International Conference on Communications Workshops (ICC Workshops)}.\hskip 1em plus 0.5em minus 0.4em\relax IEEE, 2023, pp. 1118--1123.

\bibitem{timilsina2017secure}
S.~Timilsina and G.~Amarasuriya, ``Secure communication in relay-assisted massive mimo downlink,'' in \emph{GLOBECOM 2017-2017 IEEE Global Communications Conference}.\hskip 1em plus 0.5em minus 0.4em\relax IEEE, 2017, pp. 1--7.

\bibitem{zhu2015linear}
J.~Zhu, R.~Schober, and V.~K. Bhargava, ``Linear precoding of data and artificial noise in secure massive mimo systems,'' \emph{IEEE Transactions on Wireless Communications}, vol.~15, no.~3, pp. 2245--2261, 2015.

\bibitem{kuang2020analysis}
Q.~Kuang, J.~Gong, X.~Chen, and X.~Ma, ``Analysis on computation-intensive status update in mobile edge computing,'' \emph{IEEE Transactions on Vehicular Technology}, vol.~69, no.~4, pp. 4353--4366, 2020.

\bibitem{yang2023age}
Y.~Yang, B.~Zhang, D.~Guo, R.~Xu, C.~Su, and W.~Wang, ``Age of information optimization for privacy-preserving mobile crowdsensing,'' \emph{IEEE Transactions on Emerging Topics in Computing}, vol.~12, no.~1, pp. 281--292, 2023.

\bibitem{han2021age}
R.~Han, J.~Wang, L.~Bai, J.~Liu, and J.~Choi, ``Age of information and performance analysis for uav-aided iot systems,'' \emph{IEEE Internet of Things Journal}, vol.~8, no.~19, pp. 14\,447--14\,457, 2021.

\bibitem{jiang2025optimization}
J.~Jiang, X.~Tian, Y.~Han, C.~Hu, B.~Feng, and J.~Yu, ``Optimization for task offloading and downloading in uav-assisted mec systems with aerial to aerial collaboration,'' in \emph{2025 28th International Conference on Computer Supported Cooperative Work in Design (CSCWD)}.\hskip 1em plus 0.5em minus 0.4em\relax IEEE, 2025, pp. 667--673.

\bibitem{zhang2017securing}
G.~Zhang, Q.~Wu, M.~Cui, and R.~Zhang, ``Securing uav communications via trajectory optimization,'' in \emph{GLOBECOM 2017-2017 IEEE Global Communications Conference}.\hskip 1em plus 0.5em minus 0.4em\relax IEEE, 2017, pp. 1--6.

\bibitem{khurshid2023drl}
T.~Khurshid, W.~Ahmed, M.~Rehan, R.~Ahmad, M.~M. Alam, and A.~Radwan, ``A drl strategy for optimal resource allocation along with 3d trajectory dynamics in uav-mec network,'' \emph{IEEE Access}, vol.~11, pp. 54\,664--54\,678, 2023.

\bibitem{ming2024constrained}
F.~Ming, W.~Gong, L.~Wang, and Y.~Jin, ``Constrained multi-objective optimization with deep reinforcement learning assisted operator selection,'' \emph{IEEE/CAA Journal of Automatica Sinica}, vol.~11, no.~4, pp. 919--931, 2024.

\bibitem{sun2022resource}
H.~Sun and H.~Xi, ``Resource optimization technology using genetic algorithm in uav-assisted edge computing environment,'' \emph{Journal of Robotics}, vol. 2022, no.~1, p. 3664663, 2022.

\bibitem{sallam2020improved}
K.~M. Sallam, S.~M. Elsayed, R.~K. Chakrabortty, and M.~J. Ryan, ``Improved multi-operator differential evolution algorithm for solving unconstrained problems,'' in \emph{2020 IEEE congress on evolutionary computation (CEC)}.\hskip 1em plus 0.5em minus 0.4em\relax IEEE, 2020, pp. 1--8.

\bibitem{meng2024network}
K.~Meng, K.~Han, C.~Masouros, and L.~Hanzo, ``Network-level isac: Performance analysis and optimal antenna-to-bs allocation,'' \emph{arXiv preprint arXiv:2410.06365}, 2024.

\bibitem{florescu2014probability}
I.~Florescu, \emph{Probability and stochastic processes}.\hskip 1em plus 0.5em minus 0.4em\relax John Wiley \& Sons, 2014.

\end{thebibliography}
\bibliographystyle{IEEEtran}

\end{document}